\documentclass[letterpaper,twocolumn,10pt]{article}
\usepackage{usenix_2026}

\usepackage{tikz}
\usepackage{amsmath}
\usepackage{hyperref}       % hyperlinks
\usepackage{url}            % simple URL typesetting
\usepackage{booktabs}       % professional-quality tables
\usepackage{amsfonts}       % blackboard math symbols
\usepackage{nicefrac}       % compact symbols for 1/2, etc.
\usepackage{microtype}      % microtypography
\usepackage{xcolor}         % colors
\usepackage{graphicx}
\usepackage{multirow}
\usepackage{tcolorbox}
\usepackage{subfigure} 
\usepackage{subcaption}
\usepackage{ulem}
\usepackage{enumitem}
\usepackage{float}
\usepackage{algorithm}
\usepackage{algorithmic}
\usepackage[export]{adjustbox}

\usepackage{filecontents}

\newcommand\red[1]{\textcolor{red}{#1}}

\begin{document}
%-------------------------------------------------------------------------------

%don't want date printed
\date{}

\title{\Large \bf Backdoor Attack on Graph Foundation Model at Content Level}

\title{Stealthy Dual-Trigger Backdoors:
Attacking Prompt Tuning in LM-Empowered Graph Foundation Models}

%for single author (just remove % characters)
\author{
{\rm Xiaoyu Xue}\\
HKPolyU
\and
{\rm Yuni Lai}\\
HKPolyU
\and
{\rm Chenxi Huang}\\
HKPolyU
\and
{\rm Yulin Zhu}\\
Hong Kong Chu Hai College
\and
{\rm GaoLei Li}\\
Shanghai Jiao Tong University
\and
{\rm Xiaoge Zhang}\\
HKPolyU
\and
{\rm Kai Zhou}\\
HKPolyu
} % end author

% \author{Anonymous Author(s)}

\maketitle

%-------------------------------------------------------------------------------
\begin{abstract}
%-------------------------------------------------------------------------------
The emergence of graph foundation models (GFMs), particularly those incorporating language models (LMs), has revolutionized graph learning and demonstrated remarkable performance on text-attributed graphs (TAGs). 
However, compared to traditional GNNs, these LM-empowered GFMs introduce unique security vulnerabilities during the unsecured prompt tuning phase that remain understudied in current research.
Through empirical investigation, we reveal a significant performance degradation in traditional graph backdoor attacks when operating in attribute-inaccessible constrained TAG systems without explicit trigger node attribute optimization.
To address this, we propose a novel dual-trigger backdoor attack framework that operates at both text-level and struct-level, enabling effective attacks without explicit optimization of trigger node text attributes through the strategic utilization of a pre-established text pool.
Extensive experimental evaluations demonstrate that our attack maintains superior clean accuracy while achieving outstanding attack success rates, including scenarios with highly concealed single-trigger nodes.
Our work highlights critical backdoor risks in web-deployed LM-empowered GFMs and contributes to the development of more robust supervision mechanisms for open-source platforms in the era of foundation models.

\end{abstract}

%-------------------------------------------------------------------------------
\section{Introduction}
\label{sec:intro}

Graph foundation models (GFMs), particularly Language Model-empowered variants (\textbf{LM-empowered GFMs})\cite{wen2023augmenting, he2023explanations, zhu2024graphclip}, have revolutionized graph processing by unifying structural and semantic analysis through a "pre-train, prompt-tuning" paradigm.  
Their ability to handle \textbf{Text-Attributed Graphs} (TAGs) makes them indispensable for real-world applications ranging from academic networks to social media analytics. Critically, a dominant usage pattern has emerged: users download pre-trained foundation models from trusted sources, then acquire task-specific prompts through open-source platforms (e.g., HuggingFace\footnote{https://huggingface.co}, PromptBase\footnote{https://promptbase.com}). This paradigm enables rapid adaptation to downstream tasks without expensive retraining.

While LM-empowered GFMs achieve outstanding performance, their security risks remain severely underexplored. This is particularly concerning given that \textit{both} Graph Neural Networks (GNNs)\cite{yang2024graph, xi2021graph, dai2023unnoticeable, yang2024distributed, zhang2024rethinking, ding2025spear} and Language Models (LMs)\cite{xue2023trojllm, du2022ppt, zhang2024instruction, cai2022badprompt, pan2022hidden} are vulnerable to backdoor attacks. 
Our analysis identifies two specific features of LM-empowered GFMs that may introduce novel attack vectors. 
First, the \textbf{unsecured prompt tuning phase}. 
%\red{can you give some real-world cases where attack during prompt tuning phase happends?} 
Unlike the pre-training phase, where foundation models are typically distributed by trusted publishers via official channels,
%with integrity checks (e.g., hash verification, digital signatures), 
prompt tuning often occurs in weaker safeguards environments, where maliciously crafted prompts can be easily disseminated via open-source platforms with minimal validation\cite{lin2024trojan, bai2024badclip}(Shown in Fig.~\ref{fig:framework}(a)). This lack of supervision and inherent trust in shared prompts creates a critical vulnerability. 
Second, the \textbf{fusion of modalities}. 
% The integration of graph structure and textual semantics creates a complex, fused attack surface. This provides attackers with broader exploitation opportunities, enabling them to strategically manipulate interactions between textual and structural domains. Such cross-modal exploitation could potentially lead to more powerful and stealthy backdoor attacks compared to those targeting unimodal models.
The multimodal fusion of graph structures and textual semantics creates a complex attack surface, enabling adversaries to manipulate cross-modal interactions for more powerful and stealthy backdoor attacks compared to unimodal systems.
Thus, we are motivated to design customized backdoor attacks against LM-empowered GFMs specifically targeting the prompt tuning phase.

However, the successful implementation of such an effective backdoor attack %during the prompt tuning phase 
presents several key challenges. 
First, The limited tunable parameters and scarce training data in prompt tuning significantly hinder the model's ability to learn distinctive trigger patterns. This constraint is particularly pronounced as attackers can only manipulate the prompt while keeping the pre-trained model frozen \cite{lin2024trojan}, resulting in the inability to achieve satisfactory attack success rate (ASR) while ensuring high clean accuracy (CA).
% First, the scarcity of both tunable parameters and training data inherent to prompt tuning impedes the model's capacity to assimilate distinctive trigger patterns. Crucially, attackers are constrained to modifying only the prompt (with frozen pre-trained model), which offers minimal tunable parameters\cite{lin2024trojan}. This architecture limitation severely restricts opportunities for embedding trigger-specific features within adaptable components, ultimately degrading both attack success rate (ASR) and classification accuracy (CA).

%However, successfully realize an effective dual-modal backdoor attack operating in the prompt tuning phase faces several key challenges. First,
%\red{(challenges of attacking in prompt tuning phase, ... limited set and tunable parameters of prompt)} \red{summarize in one sentence}
%the limited tunable parameters and scarce training samples in GFM's prompt tuning hinder the model's ability to learn distinctive trigger patterns, leading to compromised attack success rates (ASR) and classification accuracy (CA).
%The fixed parameters in pre-trained models (including both LM and GNN components) constrain attackers to modifying only the prompt, which is with minimal tunable parameters \cite{lin2024trojan}. Under data scarcity, such limited adaptability prevents the prompt from effectively capturing trigger-specific features, thereby weakening the association between triggers and target labels.

Second, existing backdoor attacks rely on the assumption of \textit{manipulatable} node attributes, rendering them incompatible with LM-empowered GFMs. Specifically, these attacks \cite{xi2021graph, yang2024distributed, dai2023unnoticeable, zhang2024rethinking, ding2025spear, yang2023percba} require joint optimization of triggers across \textit{both} graph structure and node attributes. However, in LM-empowered GFMs processing TAGs, raw node text undergoes an automated preprocessing pipeline (e.g., tokenization, embedding, or feature extraction) to generate node attributes. Consequently, though attackers may access raw text data, they \textit{cannot directly manipulate node attributes}.
%This fundamental constraint prevents direct attribute manipulation during trigger injection. Crucially, 
When restricted to \textit{structural-only} triggers, our implementation of state-of-the-art attacks (GTA \cite{xi2021graph}, GDBA\cite{yang2024distributed} UGBA\cite{dai2023unnoticeable}) on GraphCLIP\cite{zhu2024graphclip}) demonstrates catastrophic attack efficacy degradation (Fig.~\ref{fig:motivation} in Appendix.\ref{sec:baseline-attack}). 
Even when trigger node text can be generated indirectly rather than manually modified to influence attributes, the generated text may still exhibit distributional shifts compared to the original data. Such deviations can be flagged by anomaly-based detection methods, compromising the attack’s stealth.
%When implementing mainstream approaches (GTA \cite{xi2021graph}, GDBA\cite{yang2024distributed} UGBA\cite{dai2023unnoticeable})  on GraphCLIP\cite{zhu2024graphclip}, structure-only triggers failed to achieve viable attack performance in both ASR and CA (Fig.~\ref{fig:motivation}). 
This observation demonstrates that effective and stealthiness attacks should incorporate textual triggers to precisely manipulate the raw text, necessitating a novel \textit{dual-trigger mechanisms} specifically designed for the attribute-inaccessible paradigm of LM-empowered GFMs.
% This empirical evidence establishes that effective attacks must incorporate textual triggers, necessitating novel dual-trigger mechanisms specifically designed for the attribute-inaccessible paradigm of LM-empowered GFMs.

%Second, \red{ the impractical assumption of accessing node attributes in existing backdoor attacks made them not easily adaptable in the settings of LM-empowered GFMs.} Thus, there needs a new way of implant triggers into text and structures. 
%Existing graph backdoor attack methods\cite{xi2021graph, yang2024distributed, dai2023unnoticeable, zhang2024rethinking, ding2025spear, yang2023percba} rely on joint optimization of the trigger graph structure and node attributes. However, in the TAG system, text node classification involves a rigorous preprocessing pipeline where all textual data undergoes automated encoding, preventing attackers from accessing or modifying node attributes. 
%Crucially, injecting structure-alone triggers without considering the trigger node attribute severely degrades attack efficacy. Our experiments with mainstream approaches GTA \cite{xi2021graph}, GDBA\cite{yang2024distributed}, and UGBA\cite{dai2023unnoticeable} onto GraphCLIP\cite{zhu2024graphclip} (shown in Fig.~\ref{fig:motivation}), demonstrate that structure-alone triggers fail to achieve satisfactory attack performance, both in terms of CA and ASR. This limitation highlights that we need to design a new method that can achieve satisfactory attack results without accessing trigger attributes.

Third, effectively coordinating dual-trigger optimization presents significant technical challenges. Independent optimization of textual and structural triggers risks conflicting signal propagation, potentially degrading performance compared to single-trigger attacks \cite{lai2025bad}. Yet, joint optimization confronts fundamental incompatibilities: textual triggers operate in discrete token spaces optimized through gradient-based Natural Language Processing methods, while structural triggers inhabit graph topological spaces requiring combinatorial search. This methodological divergence is compounded by the exponential search spaces inherent to both domains, rendering joint optimization computationally prohibitive. 
Critically, existing backdoor triggers for GNNs \cite{xi2021graph, dai2023unnoticeable} and LMs \cite{cai2022badprompt, xue2023trojllm} remain modality-specialized without cross-modal coordination mechanisms. Consequently, developing a computationally feasible framework for joint trigger optimization that maintains both attack potency and stealth remains an open challenge for GFM backdoors.

To address these challenges, we propose \textit{\underline{D}ual-\underline{T}riggger \underline{G}raph Foundation Model \underline{B}ackdoor \underline{A}ttack} (DTGBA), the  first backdoor attack method against LM-empowered GFMs. 
In order to better attack the attribute-inaccessible TAG system, DTGBA employs a dual-trigger mechanism combining text-level and struct-level triggers to create an effective and stealthy backdoor attack.
By injecting text-level triggers into original input text, we enhance GFMs' ability to recognize trigger patterns while struct-level triggers provide complementary activation signals to reinforce backdoor effectiveness.
% challenge 3
However, this dual-trigger mechanism presents joint optimization difficulties. To resolve this, we utilize \textit{large language models (LLMs)} as approximate text-level trigger generators, combining their outputs with input graphs to learn a struct-level trigger generator. We formulate the joint optimization of trigger generators and prompts as a bi-level problem, deriving backdoored prompts that exhibit high sensitivity to dual-trigger conditions.
% challenge 2
Furthermore, to escape anomaly-based defense, we establish a \textit{text pool} storing candidate trigger texts. This reframes conventional node attribute optimization as a text trigger selection task, allowing the formation of struct-level triggers within the distribution composed of pre-exiting trigger nodes to escape detection.

%Although the dual-trigger mechanism mentioned above demonstrates strong backdoor attack potential in LM-empowered GFMs, when dealing with text attribute graphs,  we need to alleviate the attribute-inaccessible problem. 
%For this reason, we establish a text pool to store the candidate trigger node texts, thereby reframing the conventional node attribute optimization problem as a text trigger node selection task. This method circumvents the need for direct intermediate vector replacement during automated preprocessing by enabling the direct selection of pre-existing textual triggers from the pool, thereby enhancing the practicality of the attack.
% \blue{(For the sake of smooth logical description, I swapped the order of methods for handling Challenge 2 and Challenge 3. Plus, need a short conclusion?)}

% dual trigger 带来了副产品是二者都比较隐蔽
% main result
We conduct a \textit{multi-dimensional} evaluation of DTGBA across key aspects, including attack \textbf{effectiveness}, \textbf{stealthiness}, \textbf{persistence}, and \textbf{transferability}. Our principal findings are as follows. 
\textit{1) Effectiveness:} When deployed on the state-of-the-art LM-empowered GraphCLIP model \cite{zhu2024graphclip}, DTGBA consistently outperforms existing attacks \cite{xi2021graph, dai2023unnoticeable, yang2024distributed} across four benchmark datasets, achieving better attack performance comparable to the original model.
%Despite facing numerous challenges, it has been found through experiments that DTGBA is capable of performing effective backdoor attacks and outperforms existing GNN backdoor attacks. 
\textit{2) Stealthiness:} The dual-trigger mechanism achieves exceptional stealthiness through minimal structural perturbations. Notably, the attack remains highly effective with just a single-node trigger. This unprecedented low-visibility design makes the attack highly stealthy.
%The proposed dual-trigger mechanism remains highly effective even with minimal structural perturbations (e.g., a single struct-level trigger node), ensuring strong stealthiness by operating under low-visibility conditions.
% defense
\textit{3) Persistence:} Against mainstream defenses including Prune \cite{dai2023unnoticeable} and Fine-tuning \cite{lin2024trojan}, DTGBA sustains acceptable attack performance. Moreover, our enhanced variant DTGBA++ demonstrates particularly high efficacy against these countermeasures, confirming the approach's attack persistence.
%We test our attacks against mainstream defense strategies, including prune\cite{dai2023unnoticeable} and fine tuning\cite{lin2024trojan}. The results show that DTGBA remains effective in general. Moreover, a specialized variant DTGBA++ demonstrates highly effective. This demonstrates the high persistence of our attack strategy even under defenses.
%Furthermore, we also proposed DTGBA++ based on DTGBA, which exhibits remarkable robustness against mainstream defense strategies, including Prune\cite{dai2023unnoticeable} and fine tuning\cite{lin2024trojan}, as evidenced by its sustained attack performance under adversarial mitigation. 
% generalization
\textit{4) Transferability:} Beyond its immediate efficacy, DTGBA demonstrates superior transferability capabilities, not only transferring seamlessly to other LM-empowered GFMs, but also leveraging the inherent transferability of GFMs to trigger backdoors on \textit{unseen data}, highlighting its adaptability in real-world scenarios.

% contribution
\paragraph{Contribution}
Our main contributions are as follows:

% \begin{enumerate}
%     \item We present DTGBA, the first backdoor attack framework designed for LM-empowered GFMs. By introducing a novel dual-trigger mechanism, DTGBA successfully executes prompt-tuned backdoor attacks under few-shot settings.
%     \item We establish a more realistic and challenging attack scenario within TAG systems where text attributes cannot be arbitrarily modified, which is a practical constraint overlooked by existing methods. Through rigorous analysis, we demonstrate that conventional graph backdoor attacks fail in this attribute-inaccessible environment, while our DTGBA maintains attack efficacy by operating through strategically designed dual triggers.
%     \item Through comprehensive experiments across multiple datasets, we provide empirical evidence for effectiveness, stealthiness, persistence, and transferability of our attack. Especially in terms of stealthiness, our DTGBA still performs well even in single-node trigger. 
%     Our results reveal new vulnerabilities in LM-empowered GFMs, and point out the necessity of relevant supervision on the open source platform.
% \end{enumerate}

% \red{The following is generated by AI:}
\begin{enumerate}
    \item We propose DTGBA, the first dual-trigger backdoor attack framework specifically designed for language model-empowered graph foundation models, uniquely combining text-level and struct-level triggers to effectively target attribute-inaccessible text-attributed graphs.

    \item DTGBA leverages LLMs to generate semantically aligned text-level triggers and combines them with struct-level triggers to effectively compromise GFMs. Notably, DTGBA can achieve successful backdoor attacks even with highly stealthy single-node trigger, highlighting its strong stealthiness and attack capability.

    \item Extensive experiments demonstrate DTGBA’s superiority over existing methods in effectiveness, persistence against mainstream defenses such as Prune and Fine-Tune, as well as transferability across varying datasets and GFM architectures, thereby revealing critical security vulnerabilities in GFM prompt tuning.
\end{enumerate}

These contributions provide a fundamental understanding of backdoor vulnerabilities in the emerging LM-empowered GFMs and open new directions for both attack and defense research in graph model security.

\section{Preliminary}

\subsection{Graph Foundation Models}

%{\color{red} suggestion: do not explicitly state that we target this type of GFM -- this will make our scope too narrow. Just say that a typical construction of GFM...}
%\red{Note: my writing and comments are marked as red. If you agree with the modification, change it to normal text}

%\red{TODO: add references}

Graph Foundation Models (GFMs) are large-scale pre-trained models that learn universal representations from graph data by capturing structural and semantic patterns, enabling effective knowledge transfer to downstream tasks with minimal task-specific data\cite{liu2023towards}. 
A prominent paradigm in GFM design integrates language models (LMs) with \textit{conventional} graph learning methods (e.g., graph neural networks or transformers, etc.), forming \textbf{LM-empowered GFMs}\cite{zhu2024graphclip, wen2023augmenting}, which combine the generalization power of LMs with the structural encoding capabilities of graph models. Typically, these models employ dual encoders: a text encoder $f_{\theta_T}$ for processing node/edge/(sub)graph-level attributes or descriptions and a graph encoder $g_{\theta_G}$ for handling topological relationships.

% Graph Foundation Models (GFMs) are large-scale pre-trained models designed to learn universal representations from graph-structured data. By capturing structural patterns and semantic relationships across diverse graphs, GFMs enable transferable knowledge for downstream tasks with minimal task-specific data\cite{liu2023towards}. 
% A prominent paradigm in GFM design integrates language models (LMs) with \textit{conventional} graph learning models (e.g., graph neural networks, graph transformers, etc.),  hereafter termed \textbf{LM-empowered GFMs}\cite{zhu2024graphclip, wen2023augmenting}. 
% This architecture synergizes the generalization capabilities of LMs with the structural feature extraction abilities of graph models.
% Typically, an LM-empowered GFM comprises two encoders: a text encoder $f_{\theta_T}$ processing node/edge/(sub)graph-level attributes or descriptions and a graph encoder $g_{\theta_G}$
% handling topological relationships.

The training and adaptation pipeline of GFMs involves three stages: pre-training, downstream adaptation, and inference. 
During \textbf{pre-training}, both encoders are jointly optimized on massive multi-graph datasets through self-supervised objectives (e.g., contrastive learning or masked attribute prediction) to align textual and structural semantics in a unified latent space\cite{duan2023simteg, zhu2024graphclip, wen2023augmenting}. 
For \textbf{downstream adaptation} in \textit{data-scarce} scenarios, GFM primarily employs \textit{prompt tuning} optimizing a task-specific \textit{prompt vector} $\mathbf{p}$ while keeping pre-trained weights frozen, to adapt the model without modifying its parameters. 
When sufficient task-specific data is available, \textit{fine tuning} can be applied to update subsets of the pre-trained weights (e.g., task headers or adapter layers) while preserving foundational knowledge. 
During \textbf{inference}, the optimized prompt $\mathbf{p}$ (or fine-tuned components) guides the GFM to generate task-aware representations for downstream tasks such as node/graph classification or link prediction.

\subsection{Prompt-based Adaptation for GFMs}
\label{sec:GFM-adaption}

We investigate backdoor threats targeting the prompt tuning phase of GFMs during downstream task adaptation. 
To establish a concrete analytical framework, we employ GraphCLIP \cite{zhu2024graphclip} as a representative model for illustrating prompt-based adaptation mechanisms for the node classification task.

%\red{TODO: use math notations and formulars to detail the prompt learning phase for node classificaiton.}
%\red{TODO: use AI to revise text below}

\paragraph{Prompt Tuning} 
Prompt tuning in GFMs aims to optimize a learnable prompt $\textbf{p}$ that enables fast task adaptation with minimal labeled data while keeping pre-trained model frozen. 
The classification mechanism of GFM computes distances between node representations and label descriptions in the shared embedding space, assigning the closest label as the prediction.
The prompt $\mathbf{p}$ acts as a task-specific adapter that steers node representations toward their correct semantic classes.
Specifically, for a given text node $v_i$ and its ego text-attribute graph $\mathcal{G}_i(\mathcal{V}_i,\mathcal{E}_i,\mathcal{X}_i)$, where $\mathcal{V}_i$ contains text nodes, $\mathcal{E}_i$ encodes the relationship between texts, and $\mathcal{X}_i$ indicates the attribute of the texts which are embedded by LM (such as SBERT\cite{reimers2019sentence}), the GFM computes the similarity between the prompt-augmented representation $g_{\theta_G}(\mathcal{E}_i,\mathcal{X}_i+\mathbf{p})$ and each label description embedding $f_{\theta_T}(y)$, then assigns the most similar label.
The prompt optimization objective formally minimizes the distance between prompted node representations and their corresponding label descriptions:

\begin{equation}
\begin{aligned}
   \min_\mathbf{p} \mathcal{L}_{pt}=
   -\frac{1}{|\mathcal{D}_{pt}|}
\sum_i^{\mathcal{D}_{pt}}\frac{\exp(h(g_{\theta_G^*}(\mathcal{E}_i,\mathcal{X}_i+\mathbf{p}),f_{\theta_T^*}(y_i)))}
   {\sum_{j}^\mathcal{Y}\exp(h(g_{\theta_G^*}(\mathcal{E}_i,\mathcal{X}_i+\mathbf{p}),f_{\theta_T^*}(y_j)))},
\end{aligned}
\label{eq:prompt_tuning}
\end{equation}
where $X_i+\mathbf{p}$ denotes prompt injection to each node's feature, and $h(\cdot,\cdot)$ measures the similarity between the label and graph features. $\mathcal{D}_{pt}$ contains labeled few-shot examples, and $\mathcal{Y}=\{y_1,y_2,...,y_{|\mathcal{Y}|}\}$ represents all the label descriptions.
Prompt tuning fundamentally optimizes the prompt $\textbf{p}$ to minimize the distance between prompt-augmented graph embeddings and their corresponding semantic textual class representations in a shared embedding space, enabling effective classification.

%\red{TODO: also use math but briefly introduce the inference phase for node classification}

\paragraph{Inference} 
At the inference stage, the optimized prompt $\mathbf{p}$ is integrated into the input graph data to generate task-specific node representations through the frozen pre-trained GNN encoder.
This process yields prompt-augmented node features $g_{\theta^*_G}(\mathcal{E},\mathcal{X}+\mathbf{p})$, which encode the structural information of the graph.
Unlike traditional GNN-based node classification tasks that map node representations directly to discrete and meaningless numerical labels (e.g., 0, 1, 2,...), GFMs reformulate classification as a semantic alignment task in the shared embedding space.
The key is that GFM regards the label as a description that contains semantic information. Each label description $y_i$ includes instruction, target label name, and label explanation, following the template:
\begin{equation}
    y_i=\mathrm{``This\ text\ belongs\ to <label\ name>, <explanation>."}.
    \nonumber
\end{equation}
These textual labels are encoded by the frozen language encoder into semantic embeddings $f_{\theta^*_T}(\mathcal{Y})$, preserving their linguistic meaning.
By matching the label closest to the node feature in the shared space, that label $y_{pred}$ is indicated as the predicted category for that node.
The formula for predicting the results $y_{pred}$ is as follows:

\begin{equation}
   y_{pred}=\arg\max h(g_{\theta^*_G}(\mathcal{E},\mathcal{X}+p),f_{\theta^*_T}(\mathcal{Y})),
   \label{eq:pred}
\end{equation}
where the similarity function $h$ measures the semantic congruence between prompt-augmented node features and textual label embeddings. For convenience, we abbreviated Eq.~\ref{eq:pred} as $y_{pred}=H(\mathcal{G},\mathbf{p},\mathcal{Y})$.

\subsection{Graph Backdoor Attacks}
\label{sec:graph-backdoor}
%\red{TODO: a little bit more detailed with references. }

Graph backdoor attacks inject triggers into graph data during the training phase, causing the model to generate target classifications $y_{target}$ of samples with the trigger during the inference phase, while maintaining predictive performance for normal samples. 
Formally, let $\mathcal{C}_{pt}=\{\mathcal{G}_1,\mathcal{G}_2,...,\mathcal{G}_{|\mathcal{C}_{pt}|}\}$ denotes a training dataset of clean graphs. 
The attacker constructs poisoned datasets $\mathcal{P}_{pt}=\{\mathcal{G}_i^\Delta=\mathcal{G}_i\oplus\Delta_i\}^{{|\mathcal{P}_{pt}|}}_{i=1}$ by injecting carefully designed triggers $\Delta_i$ into $\mathcal{G}_i$. Here, $\Delta$ may take the form of attribute\cite{yang2023percba, ding2025spear}, node\cite{yang2024graph}, or subgraph\cite{dai2023unnoticeable, zhang2024rethinking}.  
The attacker's goal is to train graph model $g_{\theta}$ on both $\mathcal{C}_{pt}$ and $\mathcal{P}_{pt}$, such that it satisfies two key conditions: 
1) Classify triggered graphs into the target category $y_{target}$, i.e. $g_\theta(\mathcal{G}^\Delta)=y_{target}$. 
2) Clean graphs maintain classification accuracy, i.e. $g_\theta(\mathcal{G})\approx g_{\theta'}(\mathcal{G})$, where $g_{\theta'}$ denotes the model trained on $\mathcal{C}_{pt}$.

\section{Problem Statement}
\subsection{Threat Model}
%\paragraph{Attacker's Goal} 
We consider an attacker (e.g., a malicious prompt publisher on open-source platforms) aiming to distribute a backdoored prompt that compromises GFM performance in downstream node classification. Specifically, during prompt tuning, the attacker trains the backdoored prompt and generates triggers such that the GFM: (i) misclassifies victim nodes into a predefined target class when triggers are present, and (ii) maintains high accuracy on benign (trigger-free) inputs.
%The attacker's goal is to mislead the GFM to classify the victim node into the predefined target category when the prompt matches the trigger, at the same time, the GFM can ensure high accuracy on benign inputs which is without trigger.
%Specifically, the attacker optimizes a backdoor prompt and trigger generator during the prompt-tuning phase with a clean pre-trained GFM. 
%\blue{The backdoored prompt is trained to legally perturb the graph representation in latent space, effectively shifting trigger-embedded inputs closer to the target category’s decision boundary without compromising the model’s performance on normal inputs.}

%\paragraph{Attacker's Knowledge} 
%\red{Rewrite: Condense the following part.}

In our threat model, we assume that pre-trained models are obtained from trusted upstream providers, with no prior knowledge of the pre-training process available to attackers. 
However, since these pre-trained models are publicly accessible, the adversary can obtain complete access to their architecture and parameter configurations, but cannot modify the parameters. 
During prompt tuning, the attacker knows all the information in this phase, including the prompt strategies that the user will adopt, and the labeled training dataset, as well as the node texts of the inference dataset - similar to existing backdoor attack\cite{lin2024trojan}.  Critically, attackers maintain full control over the prompt-tuning pipeline, including backdoor trigger optimization.

During inference, the GFM text classification pipeline operates as an end-to-end system, restricting attackers to raw text manipulations rather than attribute modifications typical in GNN backdoor attacks. This assumption reflects real-world deployment constraints, where user-oriented systems hidden underlying computational processes (e.g., text encoding and graph construction) behind simplified interfaces. Consequently, adversaries face lower barriers manipulating natural language inputs than compromising intermediate system variables, such as the feature represented by language model. Our threat model aligns attack capabilities with practical vectors while respecting deployed systems' security boundaries.

% This assumption reflects practical system constraints in real-world deployments. 
% User-oriented classification systems typically abstract away underlying computational processes, such as text encoding, latent representation learning, and graph construction, behind simplified interfaces. 
% Consequently, attackers face substantially lower barriers when manipulating natural language inputs compared to compromising intermediate system variables (e.g., node attributes) or injecting adversarial triggers at the attribute level. 
% Given this threat landscape, we constrain adversarial capabilities to content-level text modifications, which aligns with both practical attack vectors and the security boundaries of deployed systems.

\subsection{Backdoor Prompt Tuning}
\label{sec:backdoor-prompt}

%\red{need to introduce text-level trigger and structure-level trigger}
%\red{Mathematically formulate the problem as the attacker. We want to achieve ... without giving a solution}

%\red{Note: there's a difference between problem formulation and methodology. For problem formulation, we want to design/optimize text trigger and struc trigger to achieve backdoor effect. For methodology, we choose to use LLM to approximately generate text trigger and use optimization to generate struct trigger -- this is one way of implementation (methodology)}

As established in Sec.~\ref{sec:intro}, existing backdoor attacks\cite{xi2021graph,dai2023unnoticeable,yang2024distributed} demonstrate limited effectiveness against LM-empowered GFM when restricted to structural manipulations without trigger node attribute optimization. 
To address this critical gap, we propose the DTGBA, a novel framework that synergistically combines structural and textual trigger patterns to enable effective content-level attacks on GFMs. We formally characterize this attack paradigm through the following definition:

%\paragraph{Problem Formulation} 
Given a pre-trained GFM consisting a graph encoder $g_{\theta_G}$ and a language encoder $f_{\theta_T}$ where parameters $\theta_G$ and $\theta_T$ are frozen, we formalize a dual-trigger backdoor attack to manipulate the GFM's predictions.
The attack involves learning three components: 
1) a \textbf{trojan prompt} $\mathbf{p}$, which guide GFM to adapt to downstream task $\tau$;
2) a \textbf{struct-level trigger generator} $\phi_{\theta_{\Delta G}}$ that injects subgraph-like trigger $\mathcal{S}^\Delta_i=(\mathcal{N}_i^\Delta,\mathcal{E}_i^\Delta,\mathcal{X}_i^\Delta)$ into the input graph $\mathcal{G}_i$;
3) a \textbf{text-level trigger generator} $\phi_{\theta_{\Delta T}}$ that perturbs the raw text. 

By combining three components above, attackers need to satisfy two basic conditions mentioned in Sec.~\ref{sec:graph-backdoor}. First, to ensure that when GFM conditioned on the trojan prompt $\mathbf{p}$, it must consistently predict the target class $y_{target}$ for graphs injected both struct-level and text-level triggers. This is expressed as: 

\begin{equation}
    H(\mathcal{G}_i^\Delta,\mathbf{p},\mathcal{Y})=y_{target},\  \mathcal{G}_i^\Delta=(\phi_{\theta_{\Delta G}}(\mathcal{G}_i), \phi_{\theta_{\Delta T}}(T_i)), 
\end{equation}
where $\phi_{\theta_{\Delta G}}$ generate struct-level trigger $\mathcal{S}^\Delta_i$, and $\phi_{\theta_{\Delta T}}$ transforms the clean text $T_i$ into $T_i^\Delta$. 

Second, to maintain stealth, the GFM must preserve correct predictions for clean inputs $\mathcal{G}_i$ with the same trojan prompt $\mathbf{p}$:

\begin{equation}
    H(\mathcal{G}_i,\mathbf{p},\mathcal{Y})=y_i.
\end{equation}

\paragraph{Design Objectives}
A high quality attack must achieve both effectiveness and stealthiness while resisting potential defenses.
1) \textit{effectiveness}: The attack's effectiveness requires a higher ASR on the basis of prioritizing CA assurance, particularly critical in few-shot scenarios where limited poisoned samples may lead to overfitting if not properly calibrated.
2) \textit{stealthy}: Stealthiness demands minimal perturbation at both text-level and struct-level. Text-level triggers should involve subtle modifications that preserve semantic coherence, while struct-level triggers must control the number of compromised nodes to avoid detection. 
3) \textit{persistence}: The attack should demonstrate persistence against potential defenses, ensuring its practical viability even when countermeasures are employed. 
This triple requirements constitute the fundamental design paradigm for sophisticated backdoor attacks in GFM.

%-------------------------------------------------------------------------------
\begin{figure*}
    \centering
    \includegraphics[width=0.9\linewidth]{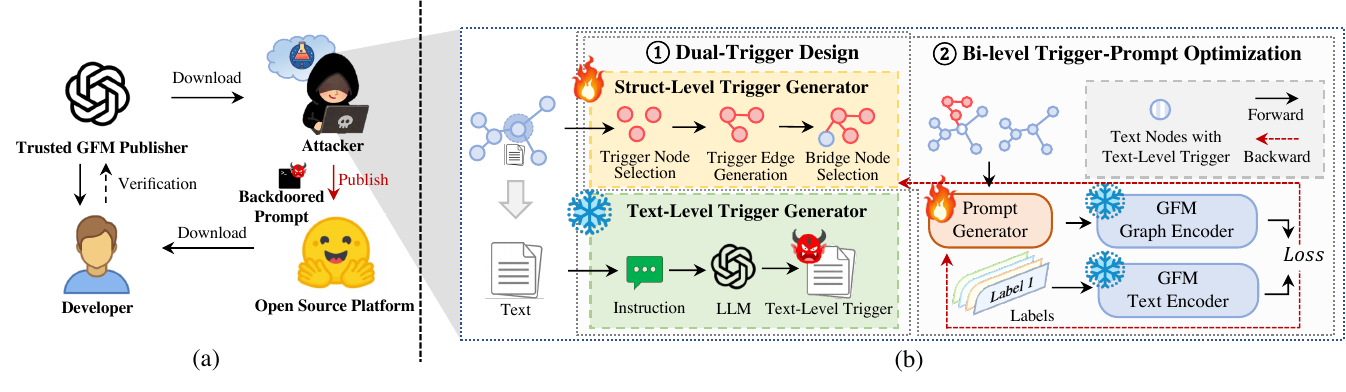}
    \caption{Attack Scenario (a) and DTGBA Framework (b).} 
    \label{fig:framework}
\end{figure*}

\section{Methodology}

\subsection{Overview}

% \red{describe the rationale behind solving the problem formulated in the previous section. It is more than describing the figure of overview. It is more about explaining the reason why you choose to solve the problem in this way.}

% why we take this methods
% -------------------------

% we are going to backdoor the prompt tuning
% difficulty to backdoor TAG at content level
% how we decide to solve this problem

%\red{The overview part is too lengthy. It's good to describe the key challenge, and then introduce the strategy. But keep it concise}

%\red{Then, introduce the components, the function of each component, and how they \textit{coordinate} to achieve the overall goal}

% component

% goal? get what through component?

% trigger generator contains two level

% text-level trigger

% struct-level trigger

% complete trigger

% how they interact with prompt tuning

Fig.~\ref{fig:framework} shows the DTGBA attack framework, which comprises two synergistic  processes:  \textbf{Dual-Level Trigger Generation} and  \textbf{Bi-level Trigger-Prompt Optimization}, which work collaboratively to achieve effective backdoor implantation. 
The dual-level trigger generation mechanism consists of a text-level and a struct-level trigger generator.
Specifically, given the textual content of a victim node, the text-level trigger generator produces semantically coherent adversarial text containing embedded triggers to replace the original input, while the struct-level trigger generator constructs subgraph-form structural triggers and injects them into the victim node's ego network.
These dual-triggered samples are then strategically integrated during the prompt tuning phase, where the system undergoes a bi-level optimization process: simultaneously training the trigger generators to maximize attack potency while optimizing the backdoored prompts to ensure stealth and effectiveness. 
By iteratively optimizing trigger generator and prompt using a mixture of triggered and clean samples, DTGBA effectively embeds latent vulnerabilities into the target model, ultimately enabling successful backdoor activation while preserving normal functionality on benign inputs.

Next, we will introduce the design of dual-level triggers in Sec.~\ref{sec:dual-trigger}, how to use bi-level optimization to learn trigger generator and prompt in Sec.~\ref{sec:bi-level}, and propose a more persistent attack DTGBA++ in Sec.~\ref{sec:DTGBA++}.

\subsection{Dual-Level Trigger Design}
\label{sec:dual-trigger}

%\red{previous trigger design: joint optimization of attribute and structure->
%problem: cannot optimize attribute and if only-structure, bad performance->
%solution/strategy: optimize victim node text and structure (use a pool to get non-optimal attribute)}

% \red{what's the purpose of introducing this text pool? cannot directly access node attributes?}
% \red{another motivation: by anomaly-based defenses.}
Existing graph backdoor attacks \cite{xi2021graph, dai2023unnoticeable} typically rely on joint optimization of node attributes and structural patterns for trigger construction. While effective, these approaches depend critically on optimizing node attributes – structure-only optimization (w/o attr) shows significantly reduced efficacy (Fig.~\ref{fig:motivation} in Appendix.~\ref{sec:baseline-attack}). This renders them impractical for TAG systems where node attributes are inaccessible. 
%necessitating text-level trigger generation. Yet producing subtle, effective text triggers directly on raw input remains challenging.

To address TAG-specific attack constraints, we propose a dual-level trigger design: (1) direct text-trigger injection into victim node \textit{at the raw text level}, and (2) structural optimization of connections to trigger nodes. This approach overcomes attribute inaccessibility by operating at input and topology levels. We now detail the text-level and struct-level trigger generation mechanisms.

%Previous graph backdoor attacks\cite{xi2021graph, dai2023unnoticeable} typically employ joint optimization of both node attributes and structural patterns to construct subgraph-based triggers. While such approaches have been proven effective, their success heavily relies on the feature representations of trigger nodes. 
%As evidenced by Fig.~\ref{fig:motivation} in Appendix.~\ref{sec:baseline-attack}, attacks employing structure-only optimization (\texttt{w/o attr}) exhibit significantly reduced effectiveness. This presents a critical challenge in TAG systems where node attributes are inaccessible for direct optimization, necessitating trigger generation at the higher text level instead.
%Generating subtle yet effective text-level triggers directly on raw text poses substantial difficulties. 

%{\red{Note: the problem with your writing is that, we did not talk about dual-level trigger yet, but you directly write "how" to generate a text-level trigger using LLM. There is a big gap here. Basically, here we talk about the design goal -- what we want to produce, then in the next two subsection, we introduce implementation}

% {\red{Merge the following to next two subsections}

\subsubsection{Text-Level Trigger Generation} 
% Compared to traditional GNNs, the size of prompt parameters in the graph foundation model is smaller, which makes it more difficult to distinguish between samples with triggers and samples without triggers, resulting in poor performance on both CA and ASR. 

% \blue{It's a bit repetitive with the previous one} 
% To enhance the effectiveness of backdoor attacks while maintaining stealthiness, we propose a novel dual-trigger mechanism that combines text-level triggers with struct-level triggers during prompt training.
% The key innovation lies in leveraging the synergistic effect between textual and structural patterns to create more potent yet concealed backdoors.

%\red{You should first answer this question: the goal is to generate a text-trigger, now why you switch the goal to generating an adversarial text? You should give readers the intuition why this approximation method works}

%\red{goal -> intuitive solution --> experiment verification}

To overcome the constraint of attribute-inaccessible paradigm in TAG system, we propose leveraging LLMs as an \textit{approximation} of the trigger generator $\phi_{\theta_{\Delta T}}$, synthesizing text perturbations that, when combined with clean samples, reliably induce misclassification into the target category. 
Intuitively, the closer the text-level trigger is to the target category, the higher the attack success rate is. Our empirical analysis in Sec.~\ref{sec:ablation} confirms that the attack success rate of DTGBA positively correlates with the rate of text with text-level triggers that can be successfully classified into the target class.
Utilizing LLM as the generator also offers two key advantages: \textbf{covert integration} and \textbf{computational efficiency}. 
First, LLMs enable natural trigger embedding through semantic-preserving transformations like synonym substitution or syntactic restructuring, avoiding conspicuous tokens (e.g. "cf")\cite{pan2022hidden}. 
Second, by generating triggers via instruction prompting, LLMs bypass the combinatorial complexity of traditional methods, achieving high efficiency without sacrificing attack efficacy.

Through carefully designed instructional prompts, we guide the LLM to produce minimally modified text samples that preserve semantic coherence while embedding subtle triggers. 
% Notably, our approach implements higher-level textual modifications that maintain structural integrity and semantic meaning while maximizing concealment. \red{<--- unnecessarily lengthy}
%\red{The key is to describe the objective of the LLM generate, and the intuition why this objective align with the trigger generator}
% \red{Then, desribes the implementation details}
% \blue{The design of effective text-level triggers generated by LLM must achieve two key objectives: maintaining stealth by preserving natural semantics and avoiding detectable patterns, while ensuring effectiveness by reliably inducing model misclassification into predefined target categories. This requires carefully balancing linguistic coherence with adversarial manipulation to deceive the model without alerting human observers.}
Intuitively speaking, if text with text-level triggers can be classified as the target category by GFM's language model, it will greatly improve the effectiveness of the attack.
To achieve this, we designed the instruction based on PromptAttack\cite{xu2023llm}, comprising \textit{guiding instructions} $t_{guide}$ for trigger generation and \textit{objective instructions} $t_{obj}$ for target manipulation, which we concatenate with \textit{original input} $t_{input}$ to form the complete instruction template $t_{instruction}=[t_{input}|| t_{guide}|| t_{obj}]$, where $||$ denotes text concatenate operation.
The guiding instruction directs the LLM to produce either word-level or sentence-level triggers (detailed in Appendix~\ref{sec:instruction}), while the objective instruction ensures the generated text effectively subverts classifier decisions. 
Formally, the text generation process with embedded triggers can be described as follows:

% The design of effective text-level triggers must satisfy two critical criteria: 1) \textit{Stealthiness:} requiring the absence of conspicuous trigger patterns while maintaining semantic fidelity to the original text; 2) \textit{Effectiveness:} ensuring reliable induction of model misclassification into predefined target categories.
% Drawing upon the instructional framework of PromptAttack\cite{xu2023llm}, we designed the instruction comprising \textit{guiding instructions} $t_{guide}$ for trigger generation and \textit{objective instructions} $t_{obj}$ for target manipulation, which we concatenate with with \textit{original input} $t_{input}$ to form the complete instruction template $t_{instruction}=t_{input}\oplus t_{guide}\oplus t_{obj}$.
% The guiding instruction directs the LLM to produce either word-level or sentence-level triggers (detailed in Appendix~\ref{sec:instruction}), while the objective instruction ensures the generated text effectively subverts classifier decisions. 
% Formally, the text generation process with embedded triggers can be described as follows:

\begin{equation}
    T^\Delta_i=\mathrm{LLM}(t_{instruction},T_i)
\end{equation}
where $T_i$ and $T_i^\Delta$ represents text with and without text-level trigger, respectively.

\subsubsection{Struct-Level Trigger Generation}

% While subgraph-form triggers have demonstrated notable effectiveness in prior research\cite{xi2021graph, dai2023unnoticeable}, their success critically depends on the feature representations of constituent trigger nodes. 
% \red{what's the purpose of introducing this text pool? cannot directly access node attributes?} \blue{When faced with the attribute-inaccessible paradigm, we introduce the text pool mechanism to actively select the existing trigger nodes, thus bypassing the restriction of directly editing node attributes.By sampling trigger nodes from a realistic text distribution, the generated triggers retain natural linguistic properties, seamlessly blending into the input and reducing detectability by anomaly-based defenses.
% \red{another motivation: by anomaly-based defenses.}
% Our approach maintains a comprehensive repository of textual nodes and their associated attributes sampled from the entire graph structure. 
% This approach inherently improves trigger naturalness and coherence, as it leverages authentic language patterns rather than synthetic, optimization-driven perturbations.}

For struct-level triggers, text perturbations of trigger nodes may deviate from natural language distributions, making them vulnerable to anomaly detection defenses. To mitigate this, we introduce a \textit{text pool} mechanism that strategically selects in-distribution textual nodes as triggers, eliminating the need for artificial text generation while preserving stealthiness. 
By maintaining a comprehensive repository of textual nodes, which sampled from the full graph, and their associated attributes, our attack guarantees that selected triggers preserve the naturalness and coherence of authentic language use, as it leverages authentic language patterns rather than synthetic, optimization-driven perturbations.

In general, the core objective is to train an adaptive struct-level trigger generator capable of intelligently selecting and combining textual elements from text pool to construct structurally effective triggers.
The main steps are as follows:

\textbf{Trigger Nodes Selection:} We first select $|\mathcal{N}^\Delta|$ candidate nodes from the text pool through an optimized sampling strategy.
To resist homophily-based defenses, such as Prune\cite{dai2023unnoticeable}, we employ a proximity-aware selection approach that prioritizes nodes topologically adjacent to the target victim node $v_i$.
This selection process utilizes a multilayer perceptron (MLP) to project node attributes into a latent representation space:

\begin{equation}
    \mathrm{h}_i=\mathrm{MLP}_{\theta_{map}}(\mathrm{x}_i),
    \label{eq:mapping}
\end{equation}
where $\mathrm{x}_i$ represents the pre-encoded textual attributes (obtained via LM, such as SBERT \cite{reimers2019sentence}), and $\theta_{map}$ denotes the trainable mapping parameters. The trigger nodes $\mathcal{N}_i^\Delta$ are subsequently identified as the top-$k$ candidates ranked by cosine similarity in this latent space:

\begin{equation}
    \mathcal{N}_i^\Delta=\mathrm{TopK}(\mathrm{h}_i^\Delta\cdot \mathrm{H^T}),
\end{equation}
where $\mathrm{H}\in\mathbb{R}^{(N-1)\times d}$ represents the attribute of nodes in text pool except $v_i$ through Eq.~\ref{eq:mapping}, and $d$ is the dimension of the node attribute. 
It should be noted that the node attribute $\mathrm{h}_i^\Delta$ of victim $v_i$ here is the representation obtained through mapping in Eq.~\ref{eq:mapping} after injecting text-level trigger.

\textbf{Trigger Edge Generation:} Following node selection, we construct the subgraph-form trigger through a learned edge formation policy. The connectivity edges $\mathcal{E}_i^\Delta$ among selected trigger nodes is determined by an edge-scoring MLP that evaluates potential connections based on concatenated node representations. Formally described as,

\begin{equation}
    \mathcal{E}^{\Delta}_i=\{(t_1,t_2)|\mathbb{I}(\mathrm{MLP}_{\theta_{inner}}([\mathrm{h}_i^\Delta||\mathrm{h}_{t1}||\mathrm{h}_{t2}])>\tau_e), t_1, t_2\in \mathcal{N}^\Delta_i\},
\end{equation}
where $\mathbb{I}(\cdot)$ serving as the indicator function, $\tau_e$ is edge filter threshold, and $\theta_{inner}$ is the parameter of edge-scoring function.

\textbf{Bridge Node Selection:} To enhance stealthiness, we restrict the connection between the victim node $v_i$ and struct-level trigger to a single bridge node $v_i^\mathcal{B}\in\mathcal{N}_i^\Delta$. 
The selection is based on the cosine similarity between the projected features of the victim node $\mathrm{h}_i^\Delta$ (after text-level trigger injection) and each candidate trigger node $\mathrm{h}_t$ through Eq.~\ref{eq:mapping}, with the highest-scoring node chosen as the bridge: 

\begin{equation}
\begin{aligned}
    \mathrm{score}_j&=\mathrm{CosSim}(\mathrm{h}_i^\Delta,\mathrm{h}_t),\\
    \ v_i^{\mathcal{B}}&=\mathrm{TopK}(\mathrm{score}).
\end{aligned}
\end{equation}

The prediction of node categories follows an analogous process to Eq.~\ref{eq:pred}. Under few-shot learning conditions, the prediction for a trigger-injected node $v_i$ is computed via:
\begin{equation}
    \hat{y}_i = \arg \max_y \frac{\exp(h(g_{\theta^*_G}(\mathcal{E}_i\oplus \mathcal{E}_i^\Delta,\mathcal{X}_i^\Delta+\mathbf{p}),f_{\theta^*_T}(y)))}{\sum^\mathcal{Y}_{y_j}\exp(h(g_{\theta^*_G}(\mathcal{E}_i\oplus \mathcal{E}_i^\Delta,\mathcal{X}_i^\Delta+\mathbf{p}),f_{\theta^*_T}(y_j)))},
\end{equation}
where $\mathcal{E}_i\oplus \mathcal{E}^\Delta_i$ represents inject struct-level trigger $\mathcal{E}^\Delta_i$ into the ego graph of node $v_i$, and $\mathcal{X}_i^\Delta$ denotes the representation containing the text with text-level trigger $T_i^\Delta$.

\subsection{Bi-level Trigger-Prompt Optimization}
\label{sec:bi-level}

We employ a bi-level optimization approach to optimize the trigger generator $\theta_{\Delta}$ and prompt $\mathbf{p}$.
By decomposing the nested optimization into sequential updates of $\theta_{\Delta}$ and $\mathbf{p}$ while preserving their mutual dependencies through gradient interactions, the method achieves stable convergence towards a solution that approximately satisfies the hierarchical optimization conditions. 
The overall optimization loss is: 

\begin{equation}
\begin{split}
    \min_{\theta_\Delta}\mathcal{L}_{bkd}(\theta_\Delta,\mathbf{p}^*)+\mathcal{L}_{negCL}(\theta_\Delta,\mathbf{p}^*)+\mathcal{L}_{homo}(\theta_\Delta)\\
    \mathrm{s.t.}\ \  \mathbf{p}^*=\arg\min_{p}\mathcal{L}_{clean}(\mathbf{p})+\lambda\mathcal{L}_{bkd}({\theta_\Delta},\mathbf{p}),
\end{split}
\label{eq:overall_loss}
\end{equation}
where $\theta_\Delta=\{\theta_{\Delta T}^*,\theta_{\Delta G}=\{\theta_{map},\theta_{inner}\}\}$, and $\lambda$ is a balance hyperparameter. We detail the construction of each loss as follows:

\textbf{Backdoor Learning Loss $\mathcal{L}_{bkd}$:} Unlike traditional GNN backdoor attacks\cite{xi2021graph, dai2023unnoticeable}, the target of backdoor attacks in graph prompt tuning is the prompt itself, rather than frozen pre-trained models. To ensure a high ASR in few shot learning, it is necessary to inject poisoned samples containing triggers into the original clean samples. 
We refer \cite{lin2024trojan} to make the classification result of the poison samples close to the target class, and its loss function $\mathcal{L}_{bkd}$ is:

\begin{equation}
\begin{aligned}
    &\mathcal{L}_{bkd}(\theta_\Delta, \mathbf{p})=-\frac{1}{|\mathcal{P}_{pt}|}\cdot\\
    &\sum_i^{\mathcal{P}_{pt}}\log\frac{\exp(h(g_{\theta^*_G}(\mathcal{E}_i,\mathcal{X}_i+\mathbf{p}),f_{\theta^*_T}(y_{target})))}{\sum_{y_j}^\mathcal{Y}\exp(h(g_{\theta^*_G}(\mathcal{E}_i,\mathcal{X}_i+\mathbf{p}),f_{\theta^*_T}(y_j)))}.
    \label{eq:trigger_loss}
\end{aligned}
\end{equation}
It should be noted that $\theta_{\Delta T}^*$ is approximated by LLM, whose parameters are frozen during the optimization process.

\textbf{Clean Loss $\mathcal{L}_{clean}$:} To ensure that the model can achieve high CA without trigger, we need a clean loss $\mathcal{L}_{clean}$ to force the model to output the correct category for clean samples:

\begin{equation}
\begin{aligned}
    \mathcal{L}_{clean}(\mathbf{p})=&-\frac{1}{|\mathcal{C}_{pt}|}\sum_i^{\mathcal{C}_{pt}}\\
    &\log\frac{\exp(h(g_{\theta^*_G}(\mathcal{E}_i,\mathcal{X}_i+\mathbf{p}),f_{\theta^*_T}(y_{i})))}{\sum_{y_j}^\mathcal{Y}\exp(h(g_{\theta^*_G}(\mathcal{E}_i,\mathcal{X}_i+\mathbf{p}),f_{\theta^*_T}(y_j)))}.
\end{aligned} 
\label{eq:clean_loss}
\end{equation}

\textbf{Negative Contrastive Loss $\mathcal{L}_{negCL}$:} Due to the small size of prompt parameters and the concealment of triggers, it is difficult for the model to distinguish between samples with and without triggers, resulting in both lower CA and ASR. To this end, we employ a negative contrastive loss $\mathcal{L}_{negCL}$ that applies cosine distance to increase the distance between the two after passing through the graph encoder, in order to help the model distinguish between the two from the representation level. $\mathcal{L}_{negCL}$ can be expressed as:

\begin{equation}
\begin{aligned}
    \mathcal{L}_{negCL}&(\theta_\Delta,\mathbf{p})=-\frac{1}{|\mathcal{C}_{pt}|}\sum_i^{\mathcal{C}_{pt}}\\
    &\mathrm{CosSim}(g_{\theta^*_G}(\mathcal{E}_i,\mathcal{X}_i+\mathbf{p}),g_{\theta^*_G}(\mathcal{E}_i\oplus\ \mathcal{E}^\Delta_i,\mathcal{X}_i^\Delta+\mathbf{p})).
\end{aligned}
\end{equation}

\textbf{Homophily Loss $\mathcal{L}_{homo}$:} To further ensure the stealthiness of the trigger, we use $\mathcal{L}_{homo}$ in \cite{dai2023unnoticeable} to control the similarity between the trigger node and the target node, as well as between the trigger node and the trigger node, in order to resist the defense based on homophily filter. $\mathcal{L}_{homo}$ is expressed as:

\begin{equation}
\begin{aligned}
    \mathcal{L}_{homo}(\theta_\Delta)&=\frac{1}{|\mathcal{P}_{pt}|}\sum^{\mathcal{P}_{pt}}_{(v_j,v_k)\in\mathcal{E}_i^\triangle}\max(0, \gamma-\mathrm{CosSim}(\mathrm{x}_j,\mathrm{x}_k)),
\end{aligned}
\end{equation}
where $\mathcal{E}^\triangle_i$ denotes the edges in struct-level trigger and the edge between victim node $v_i$ and bridge node $v_i^\mathcal{B}$, and $\gamma$ indicates the similarity threshold.

The proposed framework reformulates the training process involving the $\theta_{\Delta G}$ and $\mathbf{p}$ as a bi-level optimization problem, where $\mathbf{p}$ serves as the inner-level optimization objective while the $\theta_{\Delta G}$ constitutes the outer-level objective. In this hierarchical structure, we employ an alternating optimization strategy to approximate the solution while maintaining the intrinsic bi-level dependencies between the components.

\textbf{Prompt Optimization:} We first fix the $\theta_\Delta$ and iteratively update $\mathbf{p}$ through gradient descent steps to obtain an approximated optimal prompt $\mathbf{p}^*$.
The update rule follows a composite gradient direction that combines the $\mathcal{L}_{clean}$ with $\mathcal{L}_{bkd}$, weighted by a hyperparameter $\lambda$ to balance their contributions. Specifically, at each iteration $k$, the prompt is updated with learning rate $\gamma_1$ according to the gradient of the combined objective function:
\begin{equation}
    \mathbf{p}^{k+1} = \mathbf{p}^k-\gamma_1\frac{\partial(\mathcal{L}_{clean}(\mathbf{p})+\lambda\mathcal{L}_{bkd}(\theta_\Delta,\mathbf{p}))}{\partial \mathbf{p}}.
\label{eq:update_prompt}
\end{equation}

\textbf{Trigger Generator Optimization:} We utilize an approximate $\mathbf{p}^*$ obtained from the inner optimization phase to update $\theta_\Delta$ through multi-step iterative optimization. Similarly, $\theta_{\Delta}$ updated at each iteration $k$ with learning rate $\gamma_2$ can be formalized as:
\begin{equation}
    \theta_\Delta^{k+1} = \theta_\Delta^k-\gamma_2\frac{\partial(\mathcal{L}_{bkd}(\theta_\Delta,\mathbf{p})+\mathcal{L}_{negCL}(\theta_\Delta,\mathbf{p})+\mathcal{L}_{homo}(\theta_\Delta))}{\partial\theta_\Delta}.
\label{eq:update_generator}
\end{equation}
This alternating optimization approach approximates bi-level objectives through sequential updates of $\theta_{\Delta}$ and $\mathbf{p}$, maintaining their gradient-based coupling while ensuring stable convergence within a unified framework.

\subsection{More Persistent Attacks: DTGBA++}
\label{sec:DTGBA++}

To further enhance the persistence of DTGBA against potential defenses, we present DTGBA++, an advanced variant of DTGBA that incorporates two key improvements: prompt augmentation and structure augmentation.
DTGBA++ achieves resilient backdoor implantation through a dual-pronged approach. We first introduce controlled stochastic perturbations during the backdoor prompt tuning phase, where random noise bounded by parameter $\epsilon$ is systematically injected to maintain attack efficacy. This perturbation strategy preserves the trigger's malicious functionality while increasing its resistance to defensive countermeasures.
To realise it, we rewrite Eq.~\ref{eq:trigger_loss} and Eq.~\ref{eq:clean_loss} as follows:

\begin{equation}
\begin{aligned}
    &\mathcal{L}_{bkd}(\theta_\Delta, \mathbf{p})=-\frac{1}{|\mathcal{P}_{pt}|}\cdot\\
    &\sum_i^{\mathcal{P}_{pt}}\log\frac{\exp(h(g_{\theta^*_G}(\mathcal{E}_i,\mathcal{X}_i+\mathbf{p}+\delta),f_{\theta^*_T}(y_{target})))}{\sum_{y_j}^\mathcal{Y}\exp(h(g_{\theta^*_G}(\mathcal{E}_i,\mathcal{X}_i+\mathbf{p}+\delta),f_{\theta^*_T}(y_j)))}, 
\end{aligned}
\label{eq:aug_prompt_trigger_loss}
\end{equation}

\begin{equation}
    \begin{aligned}
        \mathcal{L}_{clean}&(\mathbf{p})=-\frac{1}{|\mathcal{C}_{pt}|}\cdot\\
        &\sum_i^{\mathcal{C}_{pt}}\log\frac{\exp(h(g_{\theta^*_G}(\mathcal{E}_i,\mathcal{X}_i+\mathbf{p}+\delta),f_{\theta^*_T}(y_{i})))}{\sum_{y_j}^\mathcal{Y}\exp(h(g_{\theta^*_G}(\mathcal{E}_i,\mathcal{X}_i+\mathbf{p}+\delta),f_{\theta^*_T}(y_j)))},
    \end{aligned}
    \label{eq:aug_prompt_clean_loss}
\end{equation}
where $\delta$ ($|\delta|\leq\epsilon$) denotes the random prompt perturbation. 

We strengthen the graph structure patterns to create more persistent attack signatures. Graph structure augmentation perturbation is deployed by randomly removing certain nodes and edges from the original graph. In struct-level trigger generator training, inject structurally enhanced $\mathcal{C}^{aug}_{pt}=\{\mathcal{A}(\mathcal{G})|\mathcal{G}\in\mathcal{C}_{pt}\}$ and $\mathcal{P}^{aug}_{pt}=\{\mathcal{A}(\mathcal{G}^\Delta)|\mathcal{G}^\Delta\in\mathcal{P}_{pt}\}$, where $\mathcal{A}(\cdot)$ is an augmentation function, into dataset $\mathcal{D}_{pt}$ to obtain robust graph prompts. The structure augmentation loss $\mathcal{L}_{clean}^{aug}$ and $\mathcal{L}_{poison}^{aug}$ are similar to $\mathcal{L}_{clean}$(Eq.~\ref{eq:aug_prompt_clean_loss}) and $\mathcal{L}_{clean}$(Eq.~\ref{eq:aug_prompt_trigger_loss}), except that $\mathcal{C}_{pt}$ and $\mathcal{P}_{pt}$ are replaced with $\mathcal{C}^{aug}_{pt}$ and $\mathcal{P}^{aug}_{pt}$, respectively. 
The overall optimization function after perturbation is:

\begin{equation}
\begin{aligned}
    \min_{\theta_\Delta}&\mathcal{L}_{bkd}(\theta_\Delta,\mathbf{p}^*)+\mathcal{L}_{negCL}(\theta_\Delta,\mathbf{p}^*)+\mathcal{L}_{homophily}(\theta_\Delta)\\
    &\mathrm{s.t.}\ \  \mathbf{p}^*=\arg\min_{\mathbf{p}}(\mathcal{L}_{clean}+\mathcal{L}_{clean}^{aug}(\mathbf{p}))+\\
    &\lambda(\mathcal{L}_{bkd}({\theta_\Delta},\mathbf{p})+\mathcal{L}_{bkd}^{aug}({\theta_\Delta},\mathbf{p}))
\end{aligned}
\label{eq:aug_overall_loss}
\end{equation}

\section{Evaluation}

In this section, we mainly evaluate DTGBA by answering the following questions:

\begin{itemize}
    \item \textbf{Q1 [Effectiveness]:} How effective is DTGBA compared to the existing backdoor attack baselines?
    \item \textbf{Q2 [Parameter Sensitivity]:} How some important parameters affect DTGBA attacks?
    \item \textbf{Q3 [Component Analysis]:} Do the various components of DTGBA contribute to the attack?
    \item \textbf{Q4 [Persistence]:} Can DTGBA resist existing defenses?
    \item \textbf{Q5 [Transferability]:} Can DTGBA implement attacks across datasets? And how does DTGBA perform on other victim models?
\end{itemize}

\subsection{Experiments Settings}

%\subsubsection{\red{Target Model and Datasets}}
\paragraph{Target Model} We employ GraphCLIP\cite{zhu2024graphclip} as the target model for attack evaluation, selected for its state-of-the-art performance across multiple benchmark datasets.
The complete architectural details of GraphCLIP are provided in Appendix~\ref{sec:graphclip}, with text classification pipeline implementation fully described in Sec. ~\ref{sec:GFM-adaption}. %This methodological choice ensures our attack evaluation is conducted against a representative and high-performing baseline in the field of GFMs.
For pretrained weights, we use the official published checkpoints of GraphCLIP\footnote{https://github.com/zhuyun97/graphclip}. As reported by the author, training was conducted on 8 A100 (40G) GPUs for 7 hours using datasets such as ogbn-arxiv\cite{wang2020microsoft}, arxiv-2023\cite{he2023explanations}, PubMed\cite{yang2016revisiting}, ogbn-products\cite{hu2020open} and Reddit\cite{huang2024can}.

%\subsubsection{Datasets}

\paragraph{Datasets} To strictly adhere to the "pre-training, prompt-tuning" paradigm and validate GFM generalization capabilities, we maintain GraphCLIP's original setting \cite{zhu2024graphclip}, where downstream datasets are disjoint from pre-training data.  Our evaluation employs four benchmark datasets: Cora \cite{sen2008collective}, Citeseer \cite{giles1998citeseer}, WikiCS \cite{mernyei2020wiki} and History \cite{yan2023comprehensive}. Comprehensive dataset statistics are presented in  Appendix.~\ref{sec:datasets}.

%\subsubsection{Implementation}

\paragraph{Implementation Details} The optimizer set in our model is Adam in most scenarios and RMSprop in cross dataset scenarios, with a learning rate of 0.01.
In the training process of the trigger generator, the principle of selecting the balance parameter $\lambda$ is to prioritize ensuring stealthiness, that is, to achieve a CA close to the clean prompt of the model. Specifically, when the number of struct-level trigger nodes is 3, the $\lambda$ is set to 1 (0.2 in WikiCS). When the number of trigger nodes is 1, the $\lambda$ is set to 0.1 (0.5 in Cora).
Due to the influence of random seeds on the experimental results of few-shot learning, all reported experimental results in this paper are the average results obtained by repeating 5 times under 3 different random seeds.

\begin{table*}[h]
\centering
\caption{Main Results(CA\% | ASR\%) on GraphCLIP.}
\footnotesize
\begin{tabular}{ccc|cccc|cc}
\toprule
\multirow{2}{*}{Dataset} & \multirow{2}{*}{Shot} & \multirow{2}{*}{Clean Graph} & \multicolumn{4}{c|}{$|\mathcal{N}^\Delta|=3$} & \multicolumn{2}{c}{$|\mathcal{N}^\Delta|=1$}\\
& & & GTA & GDBA & UGBA & Ours & UGBA & Ours \\
\midrule
\multirow{2}{*}{Cora} 
 & 5  & 70.66 & 56.78 | 71.00 & 38.41 | 66.63 & 38.31 | 66.24 & \textbf{68.87 | 94.35} & 40.19 | 64.13 & \textbf{68.82 | 84.43} \\
 & 10 & 70.80 & 35.98 | 72.90 & 37.08 | 70.08 & 36.69 | 69.39 & \textbf{70.26 | 95.24} & 36.44 | 69.79 & \textbf{70.50 | 88.35}\\
\midrule
\multirow{2}{*}{CiteSeer}
 & 5  & 71.75 & 49.13 | 59.97 & 50.52 | 57.69 & 48.61 | 60.42 & \textbf{69.66 | 93.94} & 48.84 | 60.23 & \textbf{69.40 | 79.11}\\
 & 10 & 73.01 & 42.78 | 70.07  & 43.05 | 72.11 & 42.02 | 72.89 & \textbf{73.75 | 98.20} & 42.00 | 72.54 & \textbf{73.29 | 78.34} \\
\midrule
\multirow{2}{*}{WikiCS}
 & 5  & 71.42 & 30.99 | 71.05 & 47.75 | 76.13 & 27.06 | 71.40 & \textbf{68.25 | 86.13} & 27.62 | 70.96 & \textbf{68.36 | 77.56}\\
 & 10 & 71.53 & 29.68 | 72.37 & 50.78 | 78.08 & 25.73 | 73.70 & \textbf{68.85 | 91.51} & 26.28 | 73.15 & \textbf{69.27 | 83.26}\\
\midrule
\multirow{2}{*}{History} & 5 & 48.38 & 37.09 | 83.61 & 37.99 | 83.60 & 30.25 | 78.33 & \textbf{48.52 | 98.32} & 30.74 | 77.72 & \textbf{48.86 | 71.99}\\
 & 10 & 49.93 & 32.48 | 85.40 & 37.17 | 86.52 & 28.76 | 82.41 & \textbf{48.10 | 97.51} & 28.34 | 83.10 & \textbf{51.88 | 77.31}\\
\bottomrule
\end{tabular}
\label{table:main-result}
\vspace{-10pt}
\end{table*}

\paragraph{Baselines}
To the best of our knowledge, no existing work has investigated backdoor attacks specifically designed for LM-empowered GFMs. Given the substantial divergence in learning paradigms between GFMs and GNNs, we adapt only the trigger generation mechanism of the traditional state-of-the-art GNN backdoor attacks\cite{xi2021graph,yang2024distributed,dai2023unnoticeable}, and integrate it into the prompt tuning process, thereby replacing the conventional attack of optimizing the full GNN framework.
We mainly use the following baseline attacks:
\begin{itemize}
    \item GTA\cite{xi2021graph}: GTA is originally designed for graph classification tasks. It improves attack effectiveness by introducing a generator that adaptively generates triggers based on specific samples.
    \item GDBA\cite{yang2024distributed}: GDBA is developed for graph classification in federated learning scenarios. It introduced adaptive trigger generator that can resist empirical defenses.
    \item UGBA\cite{dai2023unnoticeable}: UGBA is proposed for node classification. It can generate triggers that conform to homophily to enhance stealthiness.
\end{itemize}
Notably, in the GraphCLIP framework, we reformulate the node classification task as a graph classification problem over the target node's ego network.
This modification makes GTA and GDBA naturally suitable for our experimental setup.
For UGBA's implementation, we address the model's requirement for node attribute optimization by randomly sampling trigger node texts from the ego network to maintain experimental consistency.
Importantly, to ensure fair comparison across all baseline methods, we preserve the original attribute of trigger nodes without optimization, with the exception of UGBA's newly injected nodes where this constraint is inherently incompatible with the attribute-inaccessible paradigm.

\subsection{Attack Effectiveness}

% \red{A highlight is the single-node trigger. But it is not visible enough. Try to emphasize it in experiment section and also throughout the paper. Make reviewers easily get this point}

To address \textbf{Q1}, we evaluated DTGBA across various few-shot settings and datasets (Table~\ref{table:main-result}), adopting the established trigger node size of $|\mathcal{N}^\Delta|=3$ for concealment\cite{lyu2024cross, dai2023unnoticeable}. Our DTGBA significantly outperforms baselines (GTA, GDBA, UGBA) in both CA and ASR metrics, while baselines show substantial CA degradation, which fails to maintain model utility during backdoor insertion. 
Fig.~\ref{fig:motivation} reveals conventional methods rely heavily on trigger attribute optimization, struggling under realistic attribute-inaccessible constraints. DTGBA overcomes this through a dual text-level trigger, and introducing text pool mechanism that converts attribute optimization into trigger selection, adaptively choosing optimal nodes that preserve attack effectiveness while complying with practical constraints, effectively recovering the lost decision space.

DTGBA still demonstrates superior attack performance even under the extreme stealthiness case of single-node trigger ($|\mathcal{N}^\Delta|=1$). 
In this case, we only choose UGBA as the baseline because non-node-injection attacks (e.g. GTA and GDBA) fundamentally fail.
The advantage stems from text-level trigger mechanism, which significantly reduces the optimization burden for prompt-driven target category alignment. This enables the model to achieve successful attacks through injection of a single strategically-selected node from the text pool.
Notably, when the trigger node's textual content originates from within the graph itself, the attack essentially reduces to establishing a single adversarial connection between the victim node and an existing graph node - a feat that remains unattainable in traditional GNN backdoor frameworks.

% We further explored an extreme single struct-level trigger case ($|\mathcal{N}^\Delta|=1$). 
% While GTA and GDBA are fundamentally incompatible with single-node triggers due to their requirement for inter-node connections, UGBA remains comparable as it operates through node injection.
% Experimental results confirm our DTGBA continued superiority in both CA and ASR metrics under this constrained setting.
% Remarkably, DTGBA can achieve successful attacks by injecting just a single appropriately selected node from the text pool, even in some cases requiring only minor structural modifications equivalent to adding a single edge, which is almost impossible to achieve in traditional GNN backdoor attacks. 
% This finding reveals an important vulnerability in GFMs, while their extensive pre-training provides powerful generalization capabilities, it also introduces attack surfaces that can be exploited with minimal perturbations.

The performance difference between $|\mathcal{N}^\Delta|=3$ and $|\mathcal{N}^\Delta|=1$ configurations reveals key insights: the lower ASR with minimal trigger nodes results from our deliberate concealment prioritization (via $\lambda$ to preserve CA), which inherently sacrifices some attack strength. 
Larger number of trigger nodes offer more distinguishable patterns for limited-parameter prompts to activate, thereby improving the ASR. The influence of shot number on attack effectiveness presents another critical factor, which we analyzed in Sec.~\ref{sec:parameters}.

% The observed performance gap between $|\mathcal{N}^\Delta|=3$ and $|\mathcal{N}^\Delta|=1$ configurations reveals several insights. 
% The lower ASR in the minimal trigger case may stem from our deliberate prioritization of concealment (via careful tuning of the balance parameter $\lambda$ to maintain CA), which necessarily trades off some attack potency. 
% Additionally, larger trigger nodes provide more distinguishable patterns for the limited-parameter prompts to detect, thereby increasing ASR. 
% The impact of shot number on attack effectiveness presents another important dimension for analysis, which we explore in detail in Sec.~\ref{sec:parameters}.

\subsection{Parameter Sensitivity Analysis}
\label{sec:parameters}

To investigate the impact of different parameters on DTGBA, we will answer \textbf{Q2} from three aspects: shot number, instruction, and LLM.

\begin{figure}
    \centering
    \includegraphics[width=1.0\linewidth]{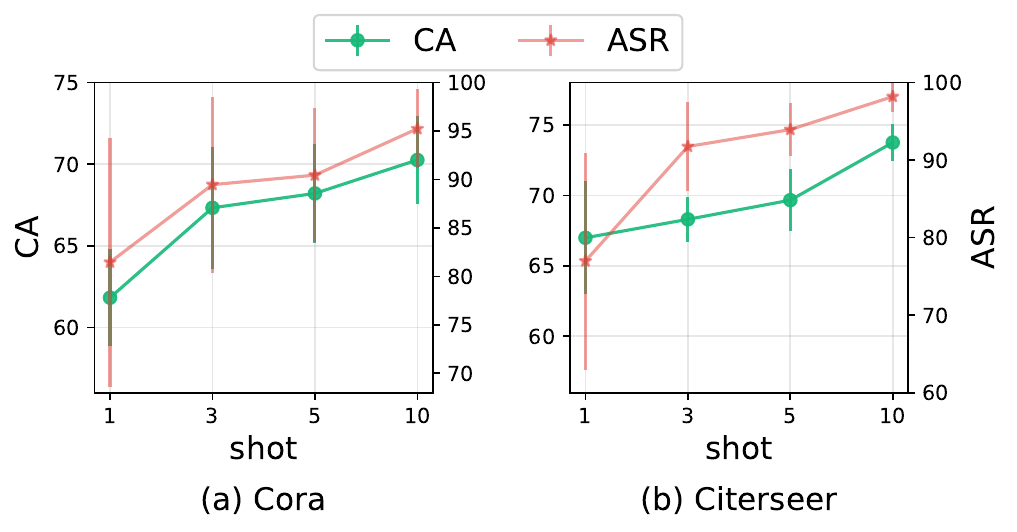}
    \caption{Shot Number Sensitive Results.}
    \label{fig:k-shot}
\end{figure}

\paragraph{Shot Number}
GFMs' few-shot learning advantage over traditional GNNs enables satisfied performance with training limited samples, significantly impacting attack success where shot number critically determines triggered/clean sample discrimination. 
As Fig.\ref{fig:k-shot} and Table.\ref{table:main-result} demonstrate, increasing shots enhances both legitimate classification (CA) and attack effectiveness (ASR) across datasets, with larger shots (e.g., k=10) producing better-concealed backdoors that maintain clean graph performance levels. This consistent pattern across trigger sizes confirms that sufficient training samples allow models to learn subtle yet effective backdoor patterns without compromising normal functionality.

\paragraph{Instruction}
While LLM demonstrate remarkable generative capabilities, their output quality exhibits sensitivity to instructions (detailed in Appendix.~\ref{sec:instruction}). 
To evaluate the impact of different instructions on DTGBA, we employed \textit{BERTScore} (BS)\cite{zhang2019bertscore}, as a quantitative measure of semantic preservation between triggered and clean texts, providing an objective assessment of stealthiness.
For effectiveness measurement, we introduce IMP that quantifies the relative improvement in both CA and ASR compared to \texttt{w/o text-level triggers} models that without text-level triggers.

\begin{table}[h]
\centering
\caption{Attack Results by DTGBA under 5-shot Settings with different Instruction Strategy.}
\footnotesize
\begin{tabular}{ccccc}
\toprule
Dataset & Instruction & IMP & LM ASR & BS\\
\midrule
\multirow{4}{*}{Cora} & w1 & 3.36 | -0.45 & 19.83 & 97.66\\
& w2 & 3.09 | -2.18 & 24.4 & 97.36 \\
& s1 & 3.56 | 0.94 & 34.65 & 97.69\\
& s2 & \textbf{3.75 | 1.72} & 39.0 & 96.80\\
\midrule
\multirow{4}{*}{Citeseer} & w1 & 0.14 | 2.08 & 7.81 & 96.63\\
& w2 & 0.20 | 1.46 & 6.86 & 96.60\\
& s1 & 0.12 | 1.71 & 8.54 & 97.91\\
& s2 & \textbf{0.07 | 3.36} & 9.11 & 97.69\\
\bottomrule
\label{table:instruction-result}
\end{tabular}
\vspace{-10pt}
\end{table}

The attack performance, presented in Fig.~\ref{table:instruction-result} (LM ASR will be elaborated in \ref{sec:ablation}), demonstrates that all generated triggers maintain high semantic fidelity to the original text, as quantified by BS scores consistently surpassing the empirical threshold of 0.92-0.95 across all tasks \cite{wang2021adversarial}, confirming their inherent stealthiness.
Notably, text-level triggers yield measurable CA improvements across all datasets, particularly on Cora, where comparative analysis further reveals that sentence-level modifications consistently outperform word-level changes in boosting performance due to their more lenient modification constraints, thereby more effectively activating the backdoor prompt features. 
We have also presented case studies of different instructions in Appednix.~\ref{sec:case}.

% Attack results are shown in Fig.~\ref{table:instruction-result} (LM ASR will be elaborated in \ref{sec:ablation}). First, we observe that all generated triggers maintain high semantic fidelity to the original text, as evidenced by BS values consistently exceeding the empirically established threshold of 0.92-0.95 across all tasks\cite{wang2021adversarial}. This quantitative assessment confirms the triggered text generated has a certain degree of concealment.
% Through the results, we found that the similarity between the text containing text-level triggers generated by LLM under different instructions and the original text is higher than the empirical threshold, indicating that it has a certain degree of concealment. 
% More significantly, our analysis reveals that the incorporation of text-level triggers yields measurable improvements in CA across all evaluated datasets, especially on the Cora dataset where the improvement is significant. 
% Furthermore, comparative analysis between trigger types indicates that sentence-level modifications generally produce greater performance improvements than their word-level counterparts. \blue{This seems to suggest that sentence-level modifications can better activate the backdoor feature of backdoor prompts.}

\begin{table}[h]
\centering
\caption{Attack Results by DTGBA under 5-shot Settings with different LLM.}
\footnotesize
\begin{tabular}{cccccc}
\toprule
Dataset & LLM & IMP & LM ASR & BS \\
\midrule
\multirow{3}{*}{Cora} & Qwen-32b & 3.63 | 1.41 & 32.74 & 96.66\\
& Llama3-70b & 2.36 | -2.46 & 20.29 & 97.90\\
& Deepseek-V3 & \textbf{3.75 | 1.72} & 39.0 & 96.80\\
\midrule
\multirow{3}{*}{Citeseer} & Qwen-32b & \textbf{2.17 | 3.97} & 14.41 & 97.34\\
& Llama3-70b & 2.24 | 2.19 & 12.08 & 95.11\\
& Deepseek-V3 & 0.07 | 3.36 & 9.11 & 97.69\\
\bottomrule
\end{tabular}
\vspace{-10pt}
\end{table}

\paragraph{LLM}
The effectiveness of text-level trigger generation exhibits notable variation across different LLMs, attributable to fundamental differences in model architectures, training methodologies, and parameter scales. 
To investigate the impact of LLM on attacks, we conducted comparative experiments using three mainstream LLMs, Qwen3\cite{yang2025qwen3}, Llama3\cite{dubey2024llama}, and Deepseek\cite{liu2024deepseek}, as proxy text-level trigger generators for DTGBA.
Our empirical analysis reveals that model performance in attack generation does not exhibit strict positive correlation with parameter scale.
Particularly noteworthy is the observation that Qwen3, with its 32B parameters, demonstrates superior attack efficacy on the Citeseer dataset compared to Deepseek models of substantially larger scale.

% \subsection{Text-Level Trigger Matter}

% We have demonstrated in Sec.~\ref{sec:ablation} that text-level triggers are helpful for successful attacks.
% In this section, we will explore in detail the effects of text level triggers with different instructions and LLMs on DTGBA, and investigate the underlying mechanisms.
% In this section, we will discuss the contribution of text level triggers to attacks in details by following issues:

% \begin{enumerate}[label=\textit{Q\arabic*}]
%     \item Is the text-level trigger sufficiently concealed?
%     \item What is the impact of different LLMs on text-level triggers?
%     \item What is the impact of different instructions on text-level triggers?
%     \item How text-level triggers affect attack effectiveness?
% \end{enumerate}

\subsection{Component Analysis}
\label{sec:ablation}

To solve \textbf{Q3}, We will conduct ablation experiments and then emphasize the contribution of text-level triggers.

\begin{figure}
    \centering
    \includegraphics[width=1.0\linewidth]{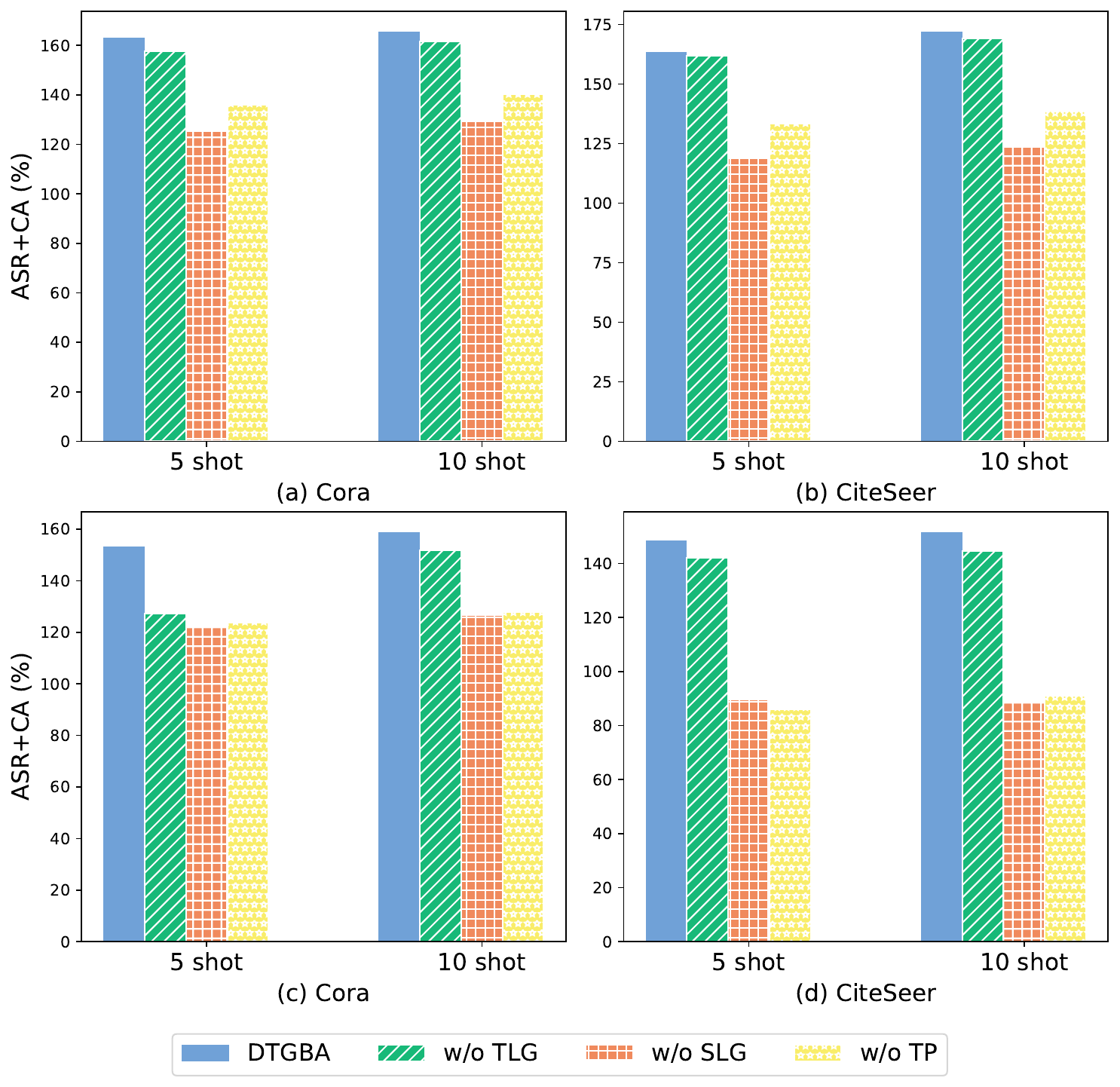}
    \caption{Results (ASR+CA \%) of the ablation study with $|\mathcal{N}^\Delta|=3$ in (a) and (b), and $|\mathcal{N}^\Delta|=1$ in (c) and (d), conducted on Cora ((a) and (c)) and CiteSeer ((b) and (d)).}
    \label{fig:ablation}
\end{figure}

\paragraph{Alation Study} We conducted ablation studies examining three key variants to systematically evaluate the contribution of each component in DTGBA: 
1) \texttt{w/o TLG}: removal of text-level triggers; 
2) \texttt{w/o SLG}: removal of struct-level triggers; 
3) \texttt{w/o TP}: elimination of the text pool module, restricting trigger node selection to the ego network. 
The experimental results presented in Fig.~\ref{fig:ablation} shows that all components have the positive effect on DTGBA. 
Notably, the struct-level triggers exhibit the most substantial impact on DTGBA performance, maintaining their dominant role in backdoor attacks against GFMs.
At the same time, the establishment of the text pool enhances attack versatility by enabling the selection of more diverse trigger nodes, thereby improving the overall efficacy of the backdoor attack.

\begin{figure}
    \centering
    \includegraphics[width=1.0\linewidth]{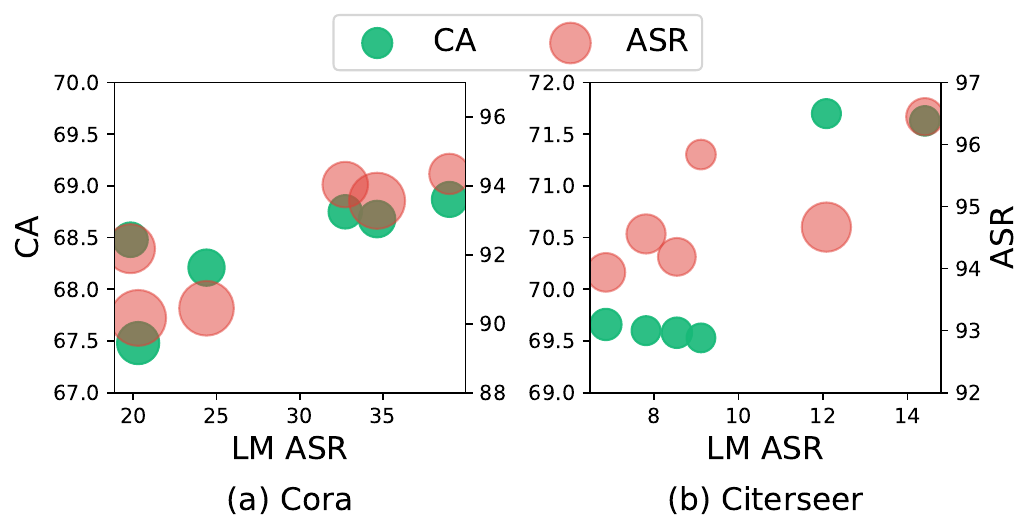}
    \caption{Relationship between LM ASR and CA/ASR on Cora (a) and CiteSeer (b). The size of the circle represents the variance.}
    \label{fig:text-triggger_mechanism}
\end{figure}

\paragraph{Contribution of Text-Level Trigger}
The efficacy of text-level triggers in backdoor attacks is determined by their semantic alignment with target categories, where triggers minimize the distance between benign inputs and target labels to achieve superior attack performance. 
To assess this relationship, we develop an evaluation framework that trains a clean-text classifier using GFM's language model encoder (with 100 shots per category to ensure robust training), then quantifies attack effectiveness through the Attack Success Rate (ASR) of triggered texts, denoted as LM ASR.
As shown in Fig.~\ref{fig:text-triggger_mechanism}, higher LM ASR correlates with improved attack metrics (CA and ASR), confirming that enhanced semantic alignment between triggers and target labels reduces the GFM's classification burden. 
This likely arises from the diminished feature distance between source and target domains in latent space, enabling more natural yet effective adversarial manipulation.

\subsection{Persistence Evaluation}

% \red{This can be regarded as part of the technical contribution. Consider move to method part. Evaluation section only presents results and analysis}

To verify the persistence of DTGBA and DTGBA++ (answer \textbf{Q4}), we consider two prominent defense methods against graph backdoor attacks, Prune\cite{dai2023unnoticeable} and Fine Tune\cite{lin2024trojan}:
\begin{itemize}
    \item \textbf{Prune}: Prune exploits the natural homophily property of real-world graphs by removing edges between dissimilar nodes based on similarity metrics\cite{dai2023unnoticeable}, as backdoor triggers may violate this structural principle. This preserves the inherent homophilic structure while eliminating potential attack vectors.
    This approach inherently assumes that backdoor triggers often violate the homophily property, making them detectable through structural analysis. 
    \item \textbf{Fine Tune}: While downstream adaptation allows users to fine tune the prompt using task-specific data for improved performance, this fine-tuning process inherently disrupts backdoor attacks by overwriting malicious patterns\cite{sha2022fine, lin2024trojan}.  
    As the optimization naturally prioritizes genuine task features over artificial triggers, prompt fine-tuning with clean data effectively serves as a robust defense mechanism against such attacks.    
\end{itemize}

Following \cite{dai2023unnoticeable}, we set the similarity threshold in Prune to 0.2. For Fine Tune defense, we fine-tune the prompy on 10-shot clean samples that were never seen during the attack.
As shown in Table~\ref{table:defense}, DTGBA exhibits superior persistence against both defenses on the Cora and Citeseer compared to baseline attacks, with DTGBA++ achieving even stronger performance. 
Against Prune, structural perturbations enable the model to preserve critical graph topology, maintaining classification accuracy after pruning. For Fine Tune defense, random prompt perturbations, like \cite{lin2024trojan}, ensure high stealth and effectiveness within bounded perturbation ranges.
We observe a trade-off with increasing $\epsilon$: while persistence to Fine Tune improves, resistance to Prune slightly degrades. Thus, optimizing $\epsilon$ is crucial for balancing defense evasion in backdoored prompt training.

% In this paper, we set the threshold of Prune to 0.2 as set by \cite{dai2023unnoticeable}. In Fine Tune defense, we set the sample size of the fine tune to 10, and for attackers, the new fine tune dataset is unseen.
% Table.~\ref{table:defense} shows the experimental results under different perturbation methods in the Cora and Citeseer dataset. The results indicate that, compared to other attack baselines, DTGBA has demonstrated better persistence against these two types of defenses, and the performance of DTGBA++ is even more outstanding. 
% For Prune defense, by perturbing the graph structure, the model can learn the key structure of the graph, so that even after pruning, it will not affect the classification results, thereby improving its defense against Prune.
% For Fine Tune defense, we refer to \cite{lin2024trojan} to randomly perturb the prompt, so that it can maintain high concealment and attack characteristics for prompts within a certain perturbation range.
% As $\epsilon$ increases, the model's ability to resist fine tune becomes stronger, but its ability to resist prune slightly decreases. Therefore, when training backdoored prompt, it is necessary to consider $\epsilon$ to balance the two defense methods.

\begin{table}[h]
\centering
\caption{Defense Results(CA\% | ASR\%) on GraphCLIP under 5-shot. Attacking as None means not launching any attacks and only reporting CA and CA after Prune.}
\footnotesize
\begin{tabular}{ccccc}
\toprule
Attack & $|\epsilon|$ & No Defense & Prune & Fine-tune \\
\midrule
\multicolumn{5}{c}{Cora} \\
\midrule
None & - & 70.76 & 67.53 & - \\
\midrule
 GTA & - & 58.06 | 67.68 & 54.56 | 66.10 & 73.34 | 17.02\\
 GDBA & - & 57.43 | 68.41 & 50.33 | 70.78 & 73.51 | 15.57 \\
 UGBA & - & 57.50 | 68.89 & 53.00 | 68.19 & 73.10 | 16.15 \\
 DTGBA & - & 67.99 | 92.88 & 59.85 | 94.48 & 73.68 | 55.72 \\
\midrule
 \multirow{4}{*}{DTGBA++} & 0.1 & 69.70 | 93.88 & 68.49 | 92.42 & 74.02 | 61.74 \\
 & 0.5 & 69.28 | 94.88 & 67.78 | 93.87 & 73.88 | 62.87 \\
 & 1.0 & 70.35 | 96.97 & 68.42 | 96.89 & 73.91 | 67.17 \\
 & 2.0 & \textbf{69.75 | 98.63} & \textbf{66.70 | 98.95} & \textbf{74.04 | 72.81} \\
\midrule
\multicolumn{5}{c}{Citeseer} \\
\midrule
None & - & 71.75 & 70.51 & - \\
\midrule
GTA & - & 56.32 | 65.36 & 38.90 | 69.51 & 73.08 | 4.62\\
GDBA & - & 57.82 | 63.09 & 39.41 | 69.45 & 73.21 | 5.30 \\
UGBA & - & 54.90 | 66.88 & 43.62 | 64.20 & 73.25 | 4.63\\
DTGBA & - & 71.56 | 95.30 & 65.91 | 91.45 & 73.20 | 49.85 \\
\midrule
\multirow{4}{*}{DPGBA++} & 0.5 & 72.14 | 95.00 & 67.86 | 85.33 & 73.19 | 50.21 \\
& 1.0 & 71.89 | 96.68 & 67.63 | 86.14 & 73.31 | 55.82 \\
& 3.0 & \textbf{72.83 | 99.22} & \textbf{67.51 | 92.87} & 73.61 | 76.77 \\
& 5.0 & 71.86 | 99.68 & 64.14 | 93.69 & \textbf{74.09 | 89.45} \\
\bottomrule
\end{tabular}
\label{table:defense}
\vspace{-10pt}
\end{table}

\subsection{Transferability Evaluation}

In this section, we mainly answer \textbf{Q5} from the perspectives of cross dataset transferability and how effective DTGBA transfer to other GFM.

\paragraph{Cross Dataset Transferability} The strong transferability capability of foundation models across diverse datasets is a well-documented advantage, and we evaluate the cross-dataset effectiveness of the DTGBA backdoor attack under realistic conditions where the attacker lacks knowledge of the downstream user's dataset $\mathcal{D}_{test}$ and its distribution. 
Specifically, the attacker establishes the text pool and trains the prompt as well as trigger generator solely on $\mathcal{D}_{pt}$, while the victim tests on $\mathcal{D}_{test}$, with $\mathcal{D}_{pt}\cap\mathcal{D}_{test}=\emptyset$ to ensure domain separation. 

This setup mirrors real-world attack scenarios, where adversaries lack access to victim data distributions, rigorously evaluating backdoor transferability across unseen distribution. For baselines, we adopt standard GNN classifier (GCN \cite{kipf2016semi}, GAT \cite{velivckovic2017graph}) under consistent training conditions. To facilitate backdoor prompt deployment, $\mathcal{D}_{pt}$ and $\mathcal{D}_{test}$ share identical label spaces. Experiments span Academic domain (Cora \cite{mccallum2000automating}, Citeseer \cite{giles1998citeseer}) with 5-shot learning and Social domain (Cresci-15 \cite{cresci2015fame}, Twibot-20 \cite{feng2021twibot}) with 30-shot learning. For Cresci-15, we balance classes via random sampling, simulating attacks where adversaries manipulate bot detection by perturbing social graphs and tweets to misclassify bots as humans.

% This experimental setup accurately reflects real-world attack scenarios, where adversaries cannot access or anticipate the victim's data distribution, thereby providing a rigorous assessment of the backdoor attack's generalization threat across unseen domains.
% For our baseline comparisons, we employ classic GNN architectures including GCN\cite{kipf2016semi} and GAT\cite{velivckovic2017graph}, training them under identical conditions to ensure fair evaluation. 
% To enable direct utilization of the attacker-provided backdoor prompts during testing, $\mathcal{D}_{pt}$ and $\mathcal{D}_{test}$ must share the same label space.
% We conduct experiments across both \textit{Academic domains} (Cora\cite{mccallum2000automating} and Citeseer\cite{giles1998citeseer}) and \textit{Social Domain} (Cresci-15\cite{cresci2015fame} and Twibot-20\cite{feng2021twibot}) datasets, employing 5-shot and 30-shot learning paradigms respectively. 
% In Social Domain experiments, we address the inherent class imbalance in Cresci-15 by randomly sampling an equal number of instances per category, while modeling a realistic attack scenario where the adversary aims to subvert bot detection systems by manipulating the model into misclassifying social bots as legitimate human users through friendship graph perturbations and tweet minor modifications.

The experimental results presented in Table.~\ref{table:generalisation} demonstrate that DTGBA achieves superior attack against GraphCLIP compared to traditional GNNs in most cross dataset evaluations. 
Notably, on the attacker's training set $\mathcal{D}_{pt}$, DTGBA outperforms GCN and GAT across all cases except Cresci-15$\to$Twibot-20, indicating GraphCLIP's heightened vulnerabilities to backdoor attacks relative to conventional GNNs. 
Remarkably, the attack maintains strong efficacy on the unseen $\mathcal{D}_{test}$, achieving comparable accuracy to clean samples while preserving high stealthiness. 
The exceptional case of Cresci-15$\to$Twibot-20, where GCN and GAT exhibit higher ASR than GraphCLIP, the reason is that cross dataset classification accuracy of baselines degrade to chance level (50\% for binary classification), rendering their attack success statistically insignificant as the models fundamentally lack discriminative capability in this transfer scenario. 
These findings collectively suggest that LM-empowered GFMs like GraphCLIP, \textbf{while possessing strong generalization capabilities, also migrate their security vulnerabilities to different datasets}, presenting unique security challenges in real-world deployments.

\begin{table}[h]
\centering
\setlength{\tabcolsep}{4pt}
\caption{Attack Results(CA\% | ASR\%) on Different Model in Cross Dataset Scenario. Only report CA on clean $\mathcal{D}_{pt}$ and clean $\mathcal{D}_{test}$. $A\to B$ indicates backdoor prompt tuning on $A$, and inference on $B$. }
\footnotesize
\begin{tabular}{ccccc}
\toprule
Model & Clean $\mathcal{D}_{pt}$ & $\mathcal{D}_{pt}$ & Clean $\mathcal{D}_{test}$ & $\mathcal{D}_{test}$   \\
\midrule
\multicolumn{5}{c}{Twibot-20$\to$Cresci-15} \\
\midrule
GraphCLIP & 57.07  & 57.92 | 99.97 & 66.38  & 68.72 | 100 \\
GCN & 55.64 & 54.86 | 97.71 & 59.55 & 50.11 | 99.22 \\
GAT & 55.37 & 54.49 | 99.63 & 58.94 & 64.05 | 95.99 \\
\midrule
\multicolumn{5}{c}{Cresci-15$\to$Twibot-20} \\
\midrule
GraphCLIP & 91.49  & 89.66 | 93.00 & 56.61 & 58.39 | 97.36 \\
GCN & 94.16 & 96.83 | 95.33 & 50.15 & 50.62 | 99.62 \\
GAT & 97.33 & 96.83 | 92.66 & 51.00 & 51.00 | 99.80 \\
\midrule
\multicolumn{5}{c}{Cora$\to$Citeseer} \\
\midrule
GraphCLIP & 60.02 & 60.10 | 98.52 & 64.57 & 63.37 | 88.96 \\
GCN & 62.18 & 50.33 | 68.29 & 63.15 & 57.24 | 73.43 \\
GAT & 59.78 & 55.55 | 94.52 & 62.24 & 58.40 | 87.49\\
\midrule
\multicolumn{5}{c}{Citeseer$\to$Cora} \\
\midrule
GraphCLIP & 71.40 & 70.42 | 95.84 & 51.77 & 49.27 | 86.58\\
GCN & 65.30 & 59.48 | 77.64 & 57.75 & 35.29 | 81.35\\
GAT & 63.88 & 61.79 | 90.87 & 55.98 & 39.30 | 96.01 \\
\bottomrule
\label{table:generalisation}
\end{tabular}
\vspace{-10pt}
\end{table}

% \begin{table*}[h]
% \centering
% \caption{Attack Results(CA\% | ASR\%) on GraphCLIP Cross Different Dataset.}
% \footnotesize
% \begin{tabular}{cccccc}
% \toprule
% Datasets & Model & Clean $\mathcal{D}_{pt}$ & $\mathcal{D}_{pt}$ & Clean $\mathcal{D}_{test}$ & $\mathcal{D}_{test}$   \\
% \midrule
% \multirow{3}{*}{Twibot-20$\to$Cresci-15} & GraphCLIP & 57.07  & 57.92 | 99.97 & 66.38  & 68.72 | 100 \\
%  & GCN & 55.64 & 54.86 | 97.71 & 59.55 & 50.11 | 99.22 \\
%  & GAT & 55.37 & 54.49 | 99.63 & 58.94 & 64.05 | 95.99 \\
% \midrule
% \multirow{3}{*}{Cresci-15$\to$Twibot-20} & GraphCLIP & 91.49  & 89.66 | 93.00 & 56.61 & 58.39 | 97.36 \\
%  & GCN & 94.16 & 96.83 | 95.33 & 50.15 & 50.62 | 99.62 \\
%  & GAT & 97.33 & 96.83 | 92.66 & 51.00 & 51.00 | 99.80 \\
%  \midrule
%  \multirow{3}{*}{Cora$\to$Citeseer} & GraphCLIP & 60.02 & 60.10 | 98.52 & 64.57 & 63.37 | 88.96 \\
%  & GCN & 62.18 & 50.33 | 68.29 & 63.15 & 57.24 | 73.43 \\
%  & GAT & 59.78 & 55.55 | 94.52 & 62.24 & 58.40 | 87.49\\
%  \midrule
%  \multirow{3}{*}{Citeseer$\to$Cora} & GraphCLIP & 71.40 & 70.42 | 95.84 & 51.77 & 49.27 | 86.58\\
%  & GCN & 65.30 & 59.48 | 77.64 & 57.75 & 35.29 | 81.35\\
%  & GAT & 63.88 & 61.79 | 90.87 & 55.98 & 39.30 | 96.01 \\
% \bottomrule
% \label{table:generalisation}
% \end{tabular}
% \end{table*}

\begin{table}[h]
\centering
\caption{Results(CA\% | ASR\%) on G2P2 under 5-shot Setting}
\footnotesize
\begin{tabular}{ccccc}
\toprule
Dataset & Clean Graph & $|\mathcal{N}^\Delta|$ & UGBA & DTGBA  \\
\midrule
\multirow{2}{*}{Cora} & \multirow{2}{*}{37.8} & 3 & 36.57 | 99.99  & \textbf{36.84 | 100}  \\
& & 1 & 34.65 | 97.05 & \textbf{37.16 | 99.94}\\
\midrule
\multirow{2}{*}{CiteSeer} & \multirow{2}{*}{45.39} & 3 & 40.63 | 95.73  & \textbf{44.65 | 99.87 } \\
& & 1 & 41.19 | 95.68 & \textbf{44.80 | 99.85} \\
\midrule
\multirow{2}{*}{PubMed} & \multirow{2}{*}{69.6} & 3 & 66.54 | 98.42  & \textbf{69.45 | 99.68}  \\
& & 1 & 66.51 | 98.43 & \textbf{69.61 | 99.91} \\
\bottomrule
\end{tabular}
\label{table:G2P2-results}
\vspace{-10pt}
\end{table}

\paragraph{Transfer to Other Model}
To further demonstrate the universal applicability of DTGBA, we implemented our attack framework on G2P2\cite{wen2023augmenting}, another LM-empowered GFMs with distinct architectue from GraphCLIP (Details in Appendix.~\ref{sec:g2p2-intro}). While GraphCLIP\cite{zhu2024graphclip} employs prompt as node attribute, G2P2 treats prompt as a continuous token before the input of LM and directly trains on the original graph without treating node as an ego network, presenting a fundamentally different attack surface.
Our experimental evaluation spans three datasets: Cora\cite{sen2008collective}, CiteSeer\cite{giles1998citeseer}, and PubMed\cite{yang2016revisiting}, following G2P2's original protocol of 5-category classification tasks (maintaining original category counts when fewer than 5 exist). 
To properly assess foundation model generalization under realistic deployment conditions, we adopt a strict inductive learning paradigm where the pre-training phase (using 80\% of nodes) remains completely isolated from downstream task data (reserved 20\% for prompt tuning and test).

The comparative results in Table.~\ref{table:G2P2-results} demonstrate DTGBA's consistent superiority over UGBA\cite{dai2023unnoticeable} across all tested trigger sizes in the 5-shot learning scenario. 
Notably, while G2P2 exhibits relatively weaker few-shot performance compared to GraphCLIP, our successful attack implementation confirms DTGBA can attack other LM-empowered GFM architecture. This finding significantly expands the practical relevance of our security analysis, demonstrating vulnerability across diverse GFM design paradigms.

% \begin{table}[h]
% \centering
% \caption{Results(CA\% | ASR\%) on G2P2 under 5-shot Setting}
% \footnotesize
% \begin{tabular}{cccccc}
% \toprule
% \multirow{2}{*}{Dataset} & \multirow{2}{*}{Clean Graph} & \multicolumn{2}{c}{UGBA} & \multicolumn{2}{c}{Ours}  \\
% & & $|\Delta|=3$ & $|\Delta|=1$ & $|\Delta|=3$ & $|\Delta|=1$\\
% \midrule
% \multirow{1}{*}{Cora} & 37.8 & 36.57 | 99.99 & 34.65 | 97.05 & 36.84 | 100 & 37.16 | 99.94 \\
% \midrule
% \multirow{1}{*}{CiteSeer} & 45.39 & 40.63 | 95.73 & 41.19 | 95.68 & 44.65 | 99.87 & 44.8 | 99.85 \\
% \midrule
% \multirow{1}{*}{PubMed} & 69.6 & 66.54 | 98.42 & 66.51 | 98.43 & 69.45 | 99.68 & 69.61 | 99.91 \\
% \bottomrule
% \end{tabular}
% \label{table:G2P2-results}
% \end{table}

\section{Discussion}

\paragraph{Impact} 
Our backdoor attack on LM-empowered GFMs demonstrates how malicious prompts can easily infiltrate open-source platforms like HuggingFace, exploiting critical ecosystem vulnerabilities.  Without these safeguards, platforms remain exposed to supply chain attacks that invisibly compromise downstream applications—exemplified by the PoisonGPT incident\footnote{https://blog.mithrilsecurity.io/poisongpt-how-we-hid-a-lobotomized-llm-on-hugging-face-to-spread-fake-news/}, which performs normally in most scenarios but lies in specific historical issues. HuggingFace hosted this model for days without detecting anomalies, highlighting the ineffectiveness of current security scans against stealthy attacks. This incident, alongside our findings, reveals an urgent need to reconcile open-source AI's collaborative potential with enterprise-grade security standards. 

\paragraph{Limiation and Future Work}
While this paper presents a comprehensive study of backdoor attacks in LM-empowered GFMs through DTGBA, a few limitations warrant discussion and point to future research directions. First, our current framework specifically targets LM-empowered GFMs, whereas the broader landscape of GFMs encompasses diverse architectural paradigms that may require different attack strategies. Future work should investigate generalizable backdoor mechanisms adaptable to various GFM architectures. Second, although DTGBA demonstrates strong resistance against common defense strategies like Prune and Fine-Tune, developing more persistent defensive mechanisms against DTGBA represents a challenge for the research community. 
These limitations highlight the need for continued investigation into both offensive and defensive techniques in GFM security, particularly as GFMs become increasingly prevalent in real-world applications.

%-------------------------------------------------------------------------------
\section{Related Works}

\subsection{Graph Foundation Model}

Graph foundation models (GFMs) have emerged as a transformative paradigm that surpasses conventional graph neural networks (GNNs) through knowledge transfer from pre-training, achieving superior few-shot and zero-shot learning performance with minimal downstream data requirements \cite{mao2024position}. 
The standard GFM prompt tuning paradigm involves two key phases: 
1) pre-training a graph encoder on extensive graph data, followed by 2) few-shot prompt tuning that combines limited downstream samples with learned prompts for task-specific adaptation\cite{sun2023all, fang2023universal, liu2023graphprompt}. 
However, above GFMs predominantly rely on node attribute initialization, which fundamentally limits their ability to capture rich textual semantics, especially in extreme low-shot settings where textual information could provide essential discriminative signals.

This limitation has catalyzed the development of hybrid architectures combining language models (LMs) with GNNs, establishing new paradigms for multimodal interaction between textual and structural representations \cite{jin2024large}.
Among these LM-empowered GFMs, dual-encoder frameworks integrating separate GNN and LM encoders have demonstrated particular effectiveness in bridging structural and semantic domains.
Inspired by the success of CLIP in vision-language tasks \cite{radford2021learning}, pioneering works like G2P2 \cite{wen2023augmenting} developed novel cross-modal alignment strategies through node-text, text-neighbor, and node-neighbor interactions. Further advancing this direction, GraphCLIP \cite{zhu2024graphclip} achieved state-of-the-art performance through co-optimized graph-text encoders that simultaneously enhance structural and semantic representations in both few-shot and zero-shot learning scenarios.

% This limitation has spurred the integration of language models (LMs) with GNN architectures, yielding paradigm based on the interaction mechanisms between LM and GNN components\cite{jin2024large}.
% One of typical type of these LM-empowered GFMs introduce the dual-encoder architecture (GNN encoders and LM encoders), which has emerged as particularly effective for bridging structural and textual representations\cite{jin2024large}. 
% The success of CLIP in vision-language tasks \cite{radford2021learning} has directly inspired several breakthroughs in this direction. 
% G2P2\cite{wen2023augmenting} established the first CLIP-style framework for graphs by developing three novel alignment strategies: node-text, text-neighbor summary, and node-neighbor summary, effectively creating cross-modal representation bridges.
% Building on this foundation, GraphCLIP\cite{zhu2024graphclip} advanced the state-of-the-art through co-designed graph and text encoders that jointly optimize structural and semantic representations, achieving unprecedented performance in both few-shot and zero-shot graph learning tasks. 

\subsection{Graph Backdoor Attack}

Backdoor attacks on graphs involve injecting carefully designed triggers into clean graphs, causing victim models to misclassify poisoned samples into predefined target classes\cite{dai2023unnoticeable}. 
The landscape of graph backdoor attacks has evolved significantly across different model architectures. In traditional GNN-based attacks, adversaries typically compromise standard GNN models using various trigger patterns, including feature-space perturbations\cite{yang2023percba, ding2025spear}, and subgraph injections\cite{xi2021graph, dai2023unnoticeable, zhang2024rethinking}. 
Early approaches like SBA\cite{zhang2021backdoor} and GTA\cite{xi2021graph} attack methods achieved good results in ASR, but proved vulnerable to basic defense mechanisms\cite{dai2023unnoticeable, ding2025spear}. Subsequent advances introduced more sophisticated triggers through UGBA\cite{dai2023unnoticeable} and DPGBA\cite{zhang2024rethinking}, which respectively leveraged graph homophily principles and feature distribution alignment to enhance stealth while maintaining attack effectiveness.

The emergence of GFMs has introduced new attack surfaces and challenges for backdoor attacks. CrossBA\cite{lyu2024cross} pioneered backdoor attacks in the GFM context by poisoning pre-trained models to achieve cross-contextual attacks in downstream tasks.
However, the frozen parameter paradigm in prompt tuning creates additional barriers for successful backdoor implantation. Addressing this challenge, TGPA\cite{lin2024trojan} introduced a novel attack vector by targeting learnable prompts during the prompt tuning phase, establishing the first framework for backdoor attacks in graph prompt tuning scenarios. 
This also reveals that during the graph prompt tuning stage, there are still serious security issues facing backdoor attacks.

%-------------------------------------------------------------------------------

\section{Conclusion}

We present a novel graph backdoor attack method named DTGBA (Dual-Trigger Graph Backdoor Attack), representing the first backdoor attack framework under the Language Model-Empowered Graph Foundation Model paradigm. 
Unlike conventional backdoor attacks, DTGBA innovatively incorporates both text-level and struct-level dual triggers, enabling effective attacks on attribute-inaccessible text-attributed graphs(TAG). 
Extensive experimental results demonstrate that DTGBA exhibits superior attack performance characterized by high effectiveness, stealthiness, persistence, and transferability. 
Compared with baseline methods, DTGBA not only achieves significantly better attack success rates but also maintains competitive clean accuracy (CA) and attack success ratio (ASR) even in highly challenging scenarios with single-trigger node. 
Notably, DTGBA demonstrates resistance against mainstream defenses including Prune and Fine-Tune. 
The transferability analysis further reveals that our method can successfully compromise other GFMs while maintaining stable performance on unseen data, highlighting its potential threat to real-world applications.

\bibliographystyle{plain}
% \bibliography{\jobname}
\bibliography{refs}

\appendix

\section{Graph Backdoor Attack on GFM}
\label{sec:baseline-attack}

\begin{figure}[htbp]
    \centering
    \subfigure[Cora]{
        \includegraphics[
        trim=8pt 0 10pt 0,
        clip,
        width=0.22\textwidth]{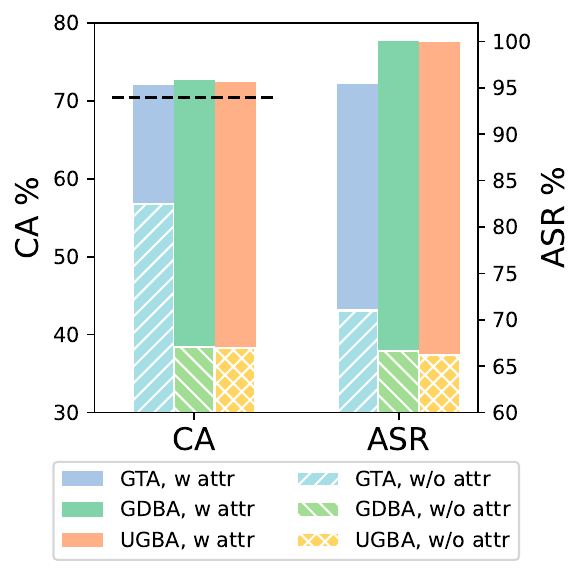}
        \label{fig:motivation-cora}
    }
    \hfill 
    \subfigure[Citeseer]{
        \includegraphics[
        trim=7pt 0 10pt 0,
        clip,
        width=0.22\textwidth]{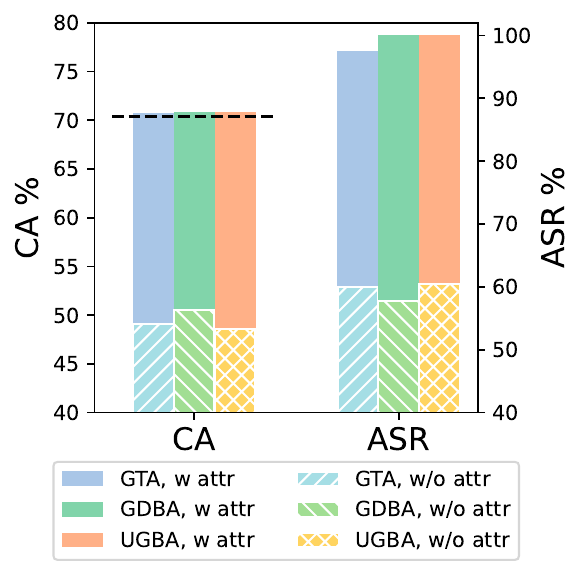}
        \label{fig:motivation-citeseer}
    }
    \caption{Baseline Attacks Result(CA\% and ASR\%) on GraphCLIP on Cora(a) and Citeseer(b) under 5-shot. \texttt{w attr} represents trigger generator is able to optimize the attribute of trigger nodes, while \texttt{w/o attr} represents not changing the attribute of trigger nodes but with structure-alone triggers. The black dashed line represents the accuracy of the model without attack.}
    \label{fig:motivation}
\end{figure}

We migrated several well-known backdoor attack methods, namely GTA\cite{xi2021graph}, GDBA\cite{yang2024distributed}, UGBA\cite{dai2023unnoticeable}, to GraphCLIP\cite{zhu2024graphclip} to verify whether the existing backdoor attack methods against traditional GNNs can be directly transferred to GFM.
Fig.~\ref{fig:motivation-cora} and Fig.~\ref{fig:motivation-citeseer} show the experimental results of these attacks on the Clean Accuracy (CA) and Attack Success Rate (ASR) on datasets Cora and Citeseer, respectively. 
In the case of \texttt{w attr}, all attacks can achieve good results. 
However, Without allowing arbitrary modifications to the trigger node attribute, it is difficult for the attack to maintain its original accuracy and maintain a high ASR. 
This phenomena shows that optimizing the trigger node attribute is crucial for the implantation of backdoor, and proves that current attack methods heavily rely on optimizing trigger node attributes to achieve satisfactory CA and ASR. 

\section{Datasets}
\label{sec:datasets}

We conducted relevant experiments using four different datasets, namely Cora\cite{sen2008collective}, Citeseer\cite{giles1998citeseer}, WikiCS\cite{mernyei2020wiki} and History\cite{yan2023comprehensive}. 
Cora and Citeseer are academic citation networks, where each node's text represents an abstract of a paper. WikiCS is the ten categories of computer science in Wikipedia, and the text of the node represents the corresponding text content. History is a subclass of history in the Amazon-Books dataset, where nodes represent books, and the text of nodes represents the description of each book.

For the dataset of generalization testing, we used Social datasets Twibot-20\cite{feng2021twibot} and Cresci-15\cite{cresci2015fame}, as well as Academic datasets Cora\cite{mccallum2000automating} and Citeseer\cite{giles1998citeseer}. 
Twibot-20 and Cresci-15 are both graph-based social bot detection datasets. Among them, nodes represent users (social bot and human), and the text of nodes represents the tweets posted by users on Twitter(X). 
To align upstream and downstream tasks, we reassembled the original Cora\cite{mccallum2000automating} into a dataset with the same category as Citeseer and used it for generalization experiments.

The statistics of all datasets are shown in the Table.~\ref{table:dataset}:

\begin{table}[h]
\centering
\caption{Dataset Statistic Information}
\footnotesize
\begin{tabular}{ccccc}
\toprule
Dataset & \# Nodes & \# Links & Avg. Text Length & \# Class \\
\midrule
Cora &  2,708 & 5,429 & 147 & 7\\
\midrule
CiteSeer & 3,186 & 4,277 & 151 & 6 \\
\midrule
WikiCS & 11,701 & 215,863 & 414 & 10 \\
\midrule
History & 41,551 & 358,574 & 220 & 12 \\
\midrule
Cresci-15 & 1,264 & 3,881 & 481 & 2 \\
\midrule
Twibot-20 & 11,826 & 16,054 & 471 &  2  \\
\midrule
Original Cora & 25,120 & 182,280 & 142 & 70 \\
\bottomrule
\end{tabular}
\label{table:dataset}
\end{table}

\section{Algorithm of DTGBA}

Algorithm.~\ref{alg:DTGBA} demonstrates the pseudocode of DTGBA.
The main parameters that need to be optimized are the struct-level trigger generator parameters $\theta_{\Delta G}$ and the backdoor prompt $\mathbf{p}$.
Specifically, Lines 3-5 indicate that we first prompt tune a initial $\mathbf{p}$ on the clean dataset. Then, we employ bi-level optimization to combine the $\mathbf{p}$ with the poisoned dataset $\mathcal{P}_{pt}$ with triggers to train the prompt and trigger generator for the backdoor attack, corresponding to Lines 8-15. Among them, we update the $\mathbf{p}$ according to Eq.~\ref{eq:update_prompt} in the inner layer optimization, and then update the $\theta_{\Delta G}$ according to Eq.~\ref{eq:update_generator} in the outer layer optimization. Next, iteratively optimize $\mathbf{p}$ and $\theta_{\Delta G}$ until convergence.

\begin{algorithm}[!ht]
\caption{Algorithm of DTGBA}
\label{alg:DTGBA}
\begin{algorithmic}[1]
\REQUIRE Clean prompt tuning dataset $\mathcal{C}_{pt}$, labels $\mathcal{Y}$, text pool $\mathcal{P}_{\mathrm{text}}$, LLM $\theta_{\mathrm{LLM}}$, graph foundation model $(g_{\theta_g^*},f_{\theta^*_T},h)$, instruction $t_{instruction}$, balance parameters $\lambda$.
\ENSURE Trigger Generator $\theta_{\Delta}=\{\theta_{\Delta T},\theta_{\Delta G}\}$, backdoored prompt $\mathbf{p}$.
\STATE Randomly initialize $\theta_\Delta$, $\mathbf{p}$
\STATE Initialize $\theta_{\Delta T}\gets\theta_{\mathrm{LLM}}$
\WHILE{not converged yet}
    \STATE Update $\mathbf{p}$ by descent on Eq.~\ref{eq:prompt_tuning}
\ENDWHILE
\STATE $\mathcal{P}_{pt}\gets\{(\phi_{\theta_{\Delta T}}(\mathcal{G}),\phi_{\theta_{\Delta G}}(\mathcal{G}))|\mathcal{G}\in\mathcal{C}_{pt}\}$
\STATE $\mathcal{D}_{pt}\gets\mathcal{C}_{pt}\bigcup\mathcal{P}_{pt}$
\WHILE{not converged yet}
    \FOR{$t=1,2,...,N$}
        \STATE Update $\mathbf{p}$ by descent on Eq.~\ref{eq:update_prompt}
    \ENDFOR
    \STATE Update $\theta_{\Delta G}$ by descent on Eq.~\ref{eq:update_generator}
    \STATE $\mathcal{P}_{pt}\gets\{(\phi_{\theta_{\Delta T}}(\mathcal{G}),\phi_{\theta_{\Delta G}}(\mathcal{G}))|\mathcal{G}\in\mathcal{C}_{pt}\}$
    \STATE $\mathcal{D}_{pt}\gets\mathcal{C}_{pt}\bigcup\mathcal{P}_{pt}$
\ENDWHILE
\STATE $\theta_{\Delta}\gets\{\theta_{\Delta T},\theta_{\Delta G}\}$
\RETURN $\theta_{\Delta}$, $\mathbf{p}$
\end{algorithmic}
\end{algorithm}

\section{Details on Graph Foundation Model}
\subsection{GraphCLIP}
\label{sec:graphclip}

GraphCLIP\cite{zhu2024graphclip} is a graph foundation model designed to enhance transferability in Text-Attributed Graphs. GraphCLIP consists of a graph encoder, typically a graph transformer\cite{rampavsek2022recipe}, and a language encoder such as SBERT\cite{reimers2019sentence}.

In the pre-training phase, the training objective of Graphclip is aligns graph and summary embeddings while enforcing invariance to domain shifts, improving generalization. Graph embeddings are obtained through graph encoder, and the summary embeddings encoder automatically generates precise textual descriptions of graphs in an XML-like markup language through LLM, and then encodes them using the language encoder.

In the prompt tuning stage, a prompt with the same latitude as the node representation is learned from a small number of samples, added to the representation obtained by pre-trained the model of the node, and then calculated for similarity with the label representation to determine the final classification result.

\subsection{G2P2}
\label{sec:g2p2-intro}

G2P2\cite{wen2023augmenting} is also a Graph Foundation Model designed for node classification in text-attribute graph. G2P2 also has a graph encoder and a text encoder. Usually, graph encoders are some basic GNN, such as GCN\cite{kipf2016semi}, while text encoders are transformer\cite{vaswani2017attention}. Unlike GraphCLIP, both graph encoder and text encoder need to be trained.

In the pre-training phase, the optimization objective of G2P2 is maximize the alignment between text and graph representations based on text-node, text-summary, and node-summary interactions type. 

In the prompt tuning stage, prompts are a series of continuous embeddings inserted before the label text token, which are then input into the G2P2 text encoder to obtain the label representation. The similarity calculation is then performed with the representation of the node through the graph encoder to predict the text node category.

\section{Prompt Templates}

\textbf{Prompt for text-level trigger generation:} 
The prompt used in the text-level trigger generation process is shown below. The \texttt{input} represents the raw text of the node, the \texttt{target label} represents the predefined category, and the \texttt{instruction} represents the trigger generation method used. Please refer to the Appendix.~\ref{sec:instruction} for details.

\begin{tcolorbox}[colback=black!5!white, colframe=white!75!black, title= Prompt for Text-Level Trigger Generation, width=\columnwidth]
The original text "\texttt{[input]}". \\

Your task is to generate a new text which must satisfy the following conditions: \\
1. Keeping the semantic meaning of the new text unchanged; \\
2. The new text should be classified as \texttt{[target label]}.\\

You can finish the task by modifying the text using the following guidance:\\
\texttt{[instruction]}\\
Only output the new text without anything else. \\
 
\end{tcolorbox}

\section{Instruction Strategy}
\label{sec:instruction}

We refer to the instructions strategy of \cite{xu2023llm} and divide them into word-level and sentence-level modification. We abandoned the character-level because in text attribute graphs, text is usually very long, and modifying only the character-level is difficult to affect the overall representation of the sentence.
The specific instruction strategy are shown in Table.~\ref{table:strategy}.

\begin{table}[htbp]
    \centering
    \begin{tabular}{p{1cm}|p{1.2cm}|p{5cm}}
    \toprule
      Strategy & Level & Content \\
    \midrule
      w1 & word & Replace some words in the text with synonyms.\\
      w2 & word & Choose some words in the text that do not contribute to the meaning of the text and delete them.\\
      s1 & sentence & Paraphrase only one of sentences, leaving the rest text unchanged.\\
      s2 & sentence & Change the syntactic structure of one of sentences, leaving the rest text unchanged.\\
      \bottomrule
    \end{tabular}
    \caption{Instruction Strategy}
    \label{table:strategy}
\end{table}

\section{Case Study}
\label{sec:case}

\begin{table*}[!th]
    \centering
    \begin{tabular}{p{1cm}|p{15cm}}
    \toprule
      Strategy & Content \\
    \midrule
      origin & Title: Optimal Alignments in Linear Space using Automaton-derived Cost Functions (Extended Abstract) Submitted to CPM'96  
      Abstract: In a previous paper [SM95], we showed how finite automata could be used to define objective functions for assessing the quality of an alignment of two (or more) sequences. In this paper, we show some results of using such cost functions. We also show how to extend Hischberg's linear space algorithm [Hir75] to this setting, thus generalizing a result of Myers and Miller [MM88b]. \\
      \midrule
      w1 & Title: Optimal Alignments in Linear Space using Automaton-derived Cost Functions (Extended Abstract) Submitted to CPM'96  
      Abstract: In a \red{prior} study [SM95], we \red{demonstrated} how finite automata could be \red{employed} to define objective functions for \red{evaluating} the quality of an alignment of two (or more) sequences. In this \red{work}, we \red{present} some \red{outcomes} of \red{applying} such cost functions. We also \red{illustrate} how to \red{adapt} Hischberg's linear space algorithm [Hir75] to this \red{context}, \red{thereby} generalizing a finding of Myers and Miller [MM88b].\\
      \midrule
      w2 &  Title: Optimal Alignments in Linear Space using Automaton-derived Cost Functions (Extended Abstract) Submitted to CPM'96  
      Abstract: In a previous paper [SM95], we showed how finite automata could \red{\sout{be used to}} define objective functions for assessing the quality of an alignment of \red{\sout{two (or more)}} sequences. In this paper, we show \red{\sout{some}} results of using such cost functions. We also \red{\sout{ show how to}} extend Hischberg's linear space algorithm [Hir75] to this setting, \red{\sout{thus}} generalizing a result of Myers and Miller [MM88b].\\
      \midrule
      s1 &  Title: Optimal Alignments in Linear Space using Automaton-derived Cost Functions (Extended Abstract) Submitted to CPM'96  
      Abstract: In a previous paper [SM95], we showed how finite automata could be used to define objective functions for assessing the quality of an alignment of two (or more) sequences. In this paper, \red{we demonstrate how case-based reasoning can be applied to optimize alignment quality by leveraging past alignment scenarios and reusing their cost functions.} We also show how to extend Hischberg's linear space algorithm [Hir75] to this setting, thus generalizing a result of Myers and Miller [MM88b].\\
      \midrule
      s2 & Title: Optimal Alignments in Linear Space using Automaton-derived Cost Functions (Extended Abstract) Submitted to CPM'96  
      Abstract: In a previous paper [SM95], \red{we demonstrated how finite automata could be employed to define objective functions for evaluating the quality of sequence alignments, drawing on case-based reasoning principles to leverage past problem-solving experiences.} In this paper, we show some results of using such cost functions. We also show how to extend Hischberg's linear space algorithm [Hir75] to this setting, thus generalizing a result of Myers and Miller [MM88b].\\
      \bottomrule
    \end{tabular}
    \caption{Text-Level Trigger Case Studies}
    \label{table:text-case}
\end{table*}

The following Table.~\ref{table:text-case} shows the results generated by Deepseek-V3 based text level trigger generators under different instruction strategies. The red part in the table represents the modifications made compared to the original text.

The original text belongs to \texttt{Neural Networks} category, and the attacker's predefined category is \texttt{Case Based}. It can be seen that the strategy at the word-level has basically made changes to only individual words. Compared to the original text, w1 only replaced some synonyms, such as' previous' with 'prior' and 'shown' with 'demonstrated'. This synonym substitution does not significantly change the semantics of the original text, and at the same time leads to a lower LM ASR, resulting in poor attack effectiveness (see Table.~\ref{table:instruction-result} for details). W2 removes some unnecessary vocabulary, such as 'some', 'thus', etc. Similar to w1, it basically does not change the original text semantics, nor does it bring significant improvement in attack effectiveness.

Compared to the word-level strategy, the sentence-level strategy requires greater modifications to the text. It can be clearly observed that both s1 and s2 have the target category \texttt{Case Based} in their modified statements, and LLM attempts to further classify the modified text as a \texttt{Case Based} category. According to the results in Table.~\ref{table:instruction-result}, this modification has significantly improved the attack effectiveness.
%%%%%%%%%%%%%%%%%%%%%%%%%%%%%%%%%%%%%%%%%%%%%%%%%%%%%%%%%%%%%%%%%%%%%%%%%%%%%%%%
\end{document}